\documentclass[%
 reprint, 
 prb,
 amsmath,amssymb,
 aps,
]{revtex4-2}

\usepackage{graphicx}
\usepackage{dcolumn}
\usepackage{bm}
\usepackage[colorlinks, allcolors=blue]{hyperref}
\usepackage[all]{hypcap}
\usepackage{blkarray}
\usepackage{amsmath}
\usepackage{svg}
\usepackage{caption}
\usepackage{comment}
\usepackage{ulem}
\usepackage{subcaption}
\usepackage{multirow}
\usepackage{rotating}
\usepackage{nicematrix}

\newcommand{\ket}[1]{\vert #1\rangle}

\newcommand{\braket}[2]{\langle #1\vert #2 \rangle}
\DeclareMathOperator{\Tr}{Tr}

\newcommand{\rrbf}{$2r$\text{-BF }}
\newcommand{\rbfeq}{{r\text{-BF}}}
\newcommand{\rjbf}{$r$\text{-JBF }}
\newcommand{\rjbfeq}{{r\text{-JBF}}}
\newcommand{\rhf}{$r$\text{-HF }}
\newcommand{\rhfeq}{{r\text{-HF}}}
\newcommand{\btheta}{{\bm{\theta}}}
\newcommand{\rjnnbfeq}{{r\text{-JNNBF}}}

\newcommand{\rjsdeq}{{r\text{-JSD}}}
\newcommand{\pref}[2]{\hyperref[#1]{\ref{#1}(#2)}}
\newcommand{\rvline}{\hspace*{-\arraycolsep}\vline\hspace*{-\arraycolsep}}

\def\urbana{
The Anthony J. Leggett Institute for Condensed Matter Theory and IQUIST and NCSA Center for Artificial Intelligence Innovation and Department of Physics, University of Illinois at Urbana-Champaign, IL 61801, USA}

\begin{document}
\title{Unifying view of fermionic neural network quantum states: From neural network backflow to hidden fermion determinant states}

\author{Zejun Liu} 
\affiliation{
\urbana}
\author{Bryan K. Clark}
\email{bkclark@illinois.edu}
\affiliation{
\urbana}
\begin{abstract}
Among the variational wave functions for Fermionic Hamiltonians, neural network backflow (NNBF) and hidden fermion determinant states (HFDS) are two prominent classes to provide accurate approximations to the ground state.  Here we develop a unifying view of fermionic neural quantum states casting them all in the framework of NNBF.   NNBF wave-functions have configuration-dependent single-particle orbitals (SPO) which are parameterized by a neural network.  We show that HFDS can be written as a NNBF wave-function with a restricted low-rank $r$ additive correction to the SPO  times a neural-network generated determinant of an  $r \times r$ matrix. 
Furthermore, we show that in NNBF wave-functions, this $r \times r$ determinant can generically be removed when $r$ is less than or equal to the number of fermions,  at the cost of further complicating the additive SPO correction increasing its rank by $r$.  We numerically and analytically compare additive SPO corrections generated by the product of two matrices with inner dimension $r$.  We find that larger $r$ wave-functions span a larger space and give  evidence that simpler and more direct updates to the SPO's tend to be more expressive and better energetically. These suggest the standard NNBF approach is preferred amongst other related choices. Finally, we uncover that the row-selection used to select single-particle orbitals allows significant sign and amplitude modulation between nearby configurations and is partially responsible for the quality of NNBF and HFDS wave-functions. 
\end{abstract}
\maketitle
\section{Introduction}

Simulating quantum many body systems is difficult often requiring various approximations to make progress.  Since the early history of quantum mechanics, variational approaches have been one of these core approximations.  In the variational approach, one starts with a parameterized class of wave-functions and searches amongst this class for the lowest energy state.  Hartree Fock, a variational search over the class of non-interacting wave-functions was amongst the earliest variational approaches for simulating fermions, and works by optimizing the parameterized single particle orbitals (SPO) which make up a  Slater Determinant~\cite{Slater}.  Even today, many of the state-of-the-art approaches for fermions and frustrated magnets build on top of this non-interacting class dressing and complementing it in various ways.  For example, a standard class of Fermionic variational wave-functions is the Slater-Jastrow form, $\Psi(c) = J(c) \Psi_{\text{SD}}(c)$ , where $\Psi_{\text{SD}}(c)$ is a non-interacting Slater Determinant and $J(c)$ is a multiplicative Jastrow factor which adds significant correlation on top of the Slater Determinant~\cite{Jastrow1955,Jastrow1994,Foulkes2001}.  The Jastrow factor $J(c)$  has historically been a strictly positive function taking the form of an exponential of one and two-body operators $J(c) =\exp[-u_1(c)+u_2(c)]$.  More recently, with advances in optimization as well as the progress in machine learning (ML) architectures which approximate a wide class of functions~\cite{GibuppseScience,Hermann2023,robledo2022fermionic,Lovato2022,Choo2020,Yoshioka2021,Di2019,Pfau2020,Hermann2020,spencer2020,Inui2021,Nomura2017,Stokes2020,Ferrari2019}, Jastrows have become more sophisticated and parameterized, spanning wave-function classes such as correlated product states, RBM, etc~\cite{Huse1988,Changlani2009,Mezzacapo2009,Hermann2020,Nomura2017,Stokes2020,Ferrari2019,needs2020variational}. 

In the context of Fermion systems, to further improve the expressibility of variational wave functions, recent work has applied ML architectures to dress the Slater Determinant part of the Slater-Jastrow wave-function.  Two such paradigms include the Neural Network Backflow (NNBF)~\cite{Ruggeri2018,Di2019,Pfau2020,chen2022simulating,Hermann2020,abrahamsen2023anti} and the hidden Fermion determinant state (HFDS)~\cite{robledo2022fermionic,Lovato2022}. NNBF replaces the static set of SPO with a configuration-dependent set which are generated by a neural network.
HFDS instead works in the paradigm of projected hidden fermions using neural networks to replace the standard Slater Determinant with a larger determinant which includes SPO's from an additional projected $r$ hidden fermions.

\begin{figure*}[t]
\centering
\includegraphics[width=\textwidth]{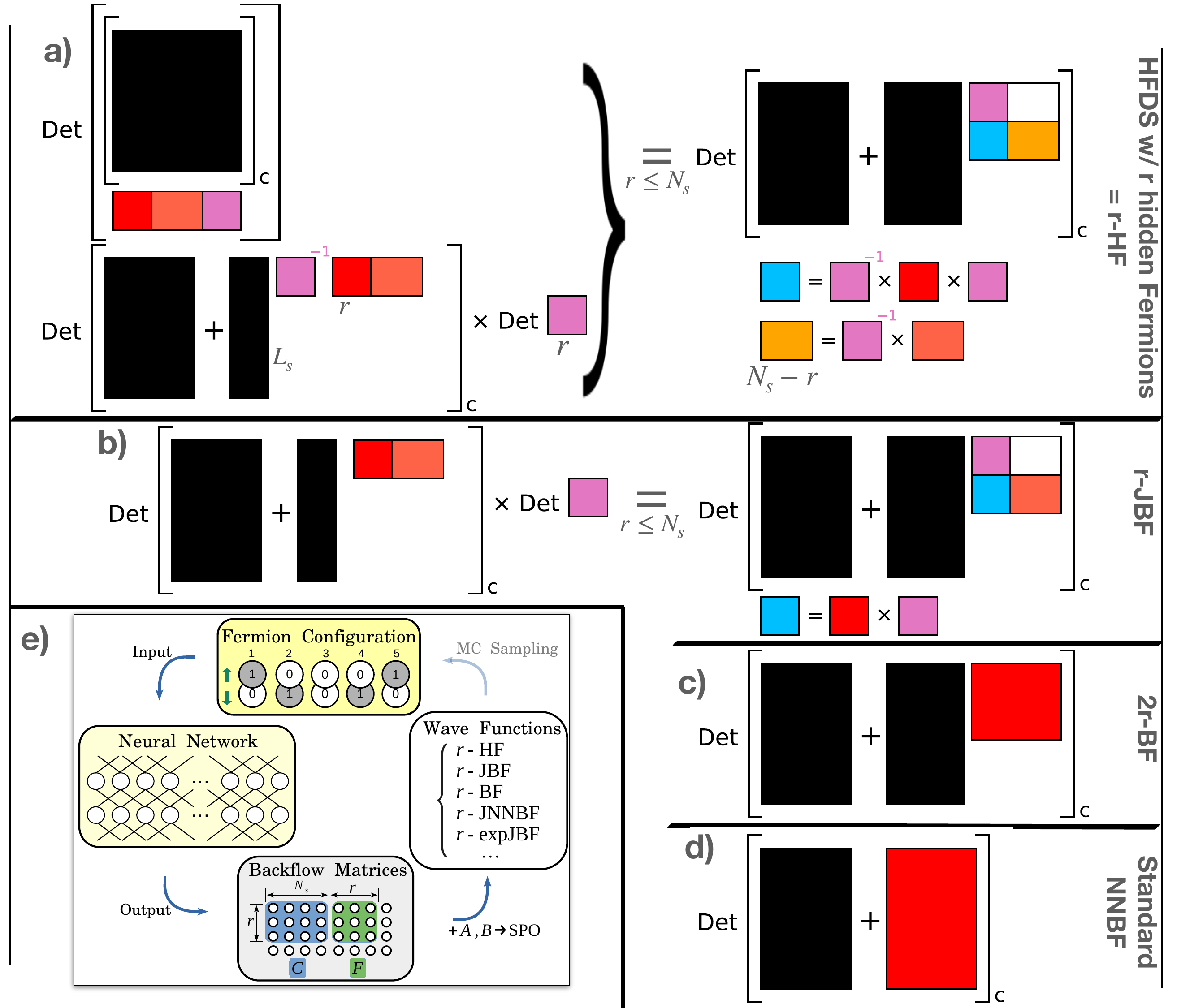}
\caption{(a-d) Visualization of wave-functions.  Red area are configuration-dependent matrices output from the NN; other colored area  depend on those outputs;  black area are optimized but configuration-independent and white area is zero.  Same color in a given line are identical (up to a trivial added identity). Brackets with $c$ in the lower-right indicate row-selection of that matrix with configuration $c$.  All wave-functions (including HFDS) can be written in the style of NNBF as an additive correction to the SPO.  Additive correction to lower wave-functions are less constrained and this additional freedom is often reflected in better energetics. Note that (c) is formally a subset of (d) even when $r>L_s$.  (b) is a subset of (c) when the NN for (c) has two additional layers and otherwise we see that both (b) and (c) and (a) and (b) can be optimized towards each other with low infidelity. The determinant factor in the l.h.s of (b) can be replaced by an exponential Jastrow with energetics of similar quality. (e) Wave function architecture for various NNBF: Fermion configuration $c$ as a set of binary numbers is input to a feed-forward neural network with two hidden layers. The output from the NN gives the $r\times N_s$-size $C_{\btheta}(c)$ and $r\times r$-size $F_{\btheta}(c)$ backflow matrices, which are then combined with the static matrices $A$ and $B$ to form SPO for various NNBF wave functions. The amplitudes from the wave functions are used to sample new fermion configurations with Monte Carlo.}
\label{fig:keyPic}
\end{figure*}
These seemingly distinct wave-functions both achieve similar significant improvements to the variational power for fermionic systems.  One cannot help but inquire about the existence of some underlying mechanism that links these states. 

In this work, we investigate explicitly the connection between NNBF and HFDS. Surprisingly, when the number of hidden fermions is less then the number of electrons, $r\leq N_s$, HFDS can be written exactly in a NNBF form as a configuration dependent update to the SPO (see Fig.~\ref{fig:keyPic}).   This set of configuration dependent SPO's can either be written as an additive low-rank $2r$ correction to the static SPO's or alternatively as a SPO correction generated from first doing a rank $r$ additive correction followed by a (right) multiplicative correction.   The way these corrections are generated by HFDS involve using the NN output in a slightly complicated fasion (i.e. taking inverse of some matrices and multiplying them by others);  we numerically demonstrate that these more complicated low-rank corrections are worse than simply doing a direct low-rank corrections and analytically show that direct low-rank corrections are always worse than the standard NNBF.    This gives evidence that the type of NNBF induced by HFDS is less expressive than the standard NNBF when the number of hidden fermions is less than the physical fermion number.  

When $r>N_s$, we can no longer view HFDS as just NNBF; instead, we have to view it as NNBF (still with an additive rank-$r$ correction to the SPO's) times an NN-generated $r\times r$ determinant.  This multiplicative factor acts like a Jastrow factor albeit one that can be negative;  toward that end, we call it a determinant factor.  We find numerically that (fixing the NNBF SPO update), using either a determinant factor,  a standard Jastrow factor or a standard Jastrow factor times an additional configuration-dependent sign given similar energetics.  When we allow the NNBF SPO update to be the standard (i.e. full rank), we find little to no energetic improvement to including any multiplicative factor.  

Finally, we investigate what feature gives NNBF (equivalently HFDS) its effective representability of low-energy ground states.  Toward that end, we rewrite a number of wave-functions as a standard (fixed, non-configuration dependent) Slater determinant times a NN-generated multiplicative factor.   We find that the NN-generated multiplicative factor coming from NNBF differs from other similar-looking wave-functions by its rapid oscillations in both sign and magnitude, a feature that we attribute to the row-selection of the determinant in these wave-functions.

\section{Neural Network Backflow}
Both the NNBF and the HFDS build on top of Slater determinants.  A Slater determinant is represented by a set of $N_s$ single particle orbitals (where $N_s$ is the total number of fermions), where each of them is scalar-valued function in space.  On lattice systems, the SPO is defined by a vector on the lattice sites.  In the continuum, it is defined by a fixed set of numbers which specify the three-dimensional (or two-dimensional) orbitals;  this may be coefficients of a Fourier expansion or points on a tri-cubic spline.  Our work will focus on the case of the lattice but the argument we make holds generically where in the continuum the neural network instead outputs the finite coefficients necessary to generate the three-dimensional orbitals.  

On a discrete lattice the SPO's are a $L_s \times N_s$ matrix $A$ where $L_s=(2s+1)L$ and $L$ is the number of lattice sites or spatial orbitals and $s$ is the spin of fermions.   Let $A[c]$ be the $N_s \times N_s$ matrix generated by taking the rows of $A$ which correspond to the locations of fermions in a configuration $c$ (equivalently evaluating the SPO function with respect to the fermion coordinates in the continuum).  The amplitude of a Slater determinant wave-function for $c$ is then $\Psi_{\text{SD}}(c)=\det A[c]$.

In NNBF, we instead make the SPO configuration-dependent,
\begin{equation}
\Psi(c) = J(c) \det (\widetilde{A}_\btheta(c)[c])
\end{equation}
where the subscript $\btheta$ are parameters to the neural network (NN) which takes as input a configuration $c$ and outputs a matrix (or a series of matrices) which is used to construct $\widetilde{A}_\btheta(c)$.  We will see examples where $\widetilde{A}_\theta(c)$  is built from a product or sum of other matrices and then the neural network outputs the configuration-dependent matrices from which the product or sum is taken.   Take $J(c)=1$ when it is not specified. 

The NN needs to be permutation invariant with respect to the input; in standard NNBF, this network is a feed forward neural network (FFNN) which takes the input $c$ as an $L_s$ bit binary number with 1's at the sites the fermions occupy.
To make comparisons fair, throughout this paper we fix all the layers (number of layers and layer widths $w$) except for a final linear layer whose output changes to be the full set of relevant matrices (In Sec.~\ref{sec:jastrow_det_benefit}, we will discuss the possibility of reconstructing the SPO for one wave-function with another by adding a few more layers, otherwise, we assume the NN's have identical architecture). 

There are various ways the new SPO can be generated from the output of the NN.  For example, through an additive correction 
\begin{equation}
 \widetilde{A}_\theta(c) =  A + M_\btheta(c)  \tag{NNBF} \label{eq:NNBF0}
\end{equation}
where $M_\btheta(c)$ is an  $L_s \times N_s$ matrix.
Alternatively, a left rotation, $\widetilde{A}_\btheta(c) = L_\btheta(c) A$ (shown in Ref.~\cite{Di2019} to be nearly identical to an additive correction) or a right rotation $\widetilde{A}_\btheta(c)=A \widetilde{F}_\btheta(c)$ can be used to generate the new SPO.  Left and right rotations are qualitatively distinct with left rotations mixing sites within each individual single particle orbital and right rotations mixing SPO between each other at fixed site.  We can even compose changes to the SPO doing first an additive correction and then a right rotation, 
\begin{equation}
\widetilde{A}_\btheta(c)=(A +M_\btheta(c)) \widetilde{F}_\btheta(c) \label{eq:Rotate} \tag{T}
\end{equation}
 
 Interestingly, right rotations can be factored out of the Slater determinant and treated as a determinant factor of the form 
\begin{equation}
J_{\det}(c)= \det \widetilde{F}_\btheta(c).  
\end{equation}
The converse of this is also true: any determinant factor $\det F_\btheta(c)$ for any $F_\btheta(c)$ whose size $r\leq N_s$ can be absorbed into Eqn.~\eqref{eq:Rotate} by letting $\widetilde{F}$ be $F$  padded with the identity up to size $N_s$. This will lead us to interchangably treat a determinant factor and a final right rotation of the SPO as equivalent (paying attention when necessary that $r\leq N_s$) throughout.  For example,  the wave-function of Eqn.~\eqref{eq:Rotate} will be JNNBF, 
\begin{equation}
    \Psi_\textrm{TNNBF} \underset{\mathrm{r\leq N_s}}{=} \Psi_\textrm{JNNBF} \equiv J_{\det} \Psi_{\textrm{NNBF}}
\end{equation}
The NN parameters $\btheta$ and $A$ are all optimized when
applied to the ground state approximation for Fermionic
Hamiltonian.  Later we will consider additional static matrices (e.g. B) which like $A$ don't depend on the neural network but whose coefficients are optimized.  

\section{Hidden Fermion Determinant States}

In HFDS, starting with an $L_s \times N_s$ set of SPO's $A$, the HFDS supplements the physical system with the introduction of $r$ hidden fermions and consequently an additional $r$ single particle orbitals as well as some number of new sites.  The $r$ additional single particle orbitals on the original `real' sites are represented by a static $L_s \times r$ matrix $B$ and $B[c]$ represents the rows selected from $B$ corresponding to the real Fermion configuration.

The evaluation of both the `real' and 'hidden' single particle orbitals on the extra sites is done implicitly. 
The NN outputs the value of the original $N_s$ and extra $r$ SPO's with respect to the (never explicitly specified) locations of the $r$ hidden fermions on the hidden sites.  This gives the $r \times N_s$ matrix $C_\btheta(c)$ for the original SPO's and the $r \times r$ matrix $F_\btheta(c)$ for the new SPO's. 

This will then give an amplitude, for $r$ hidden fermions of 
\begin{equation}
\begin{split}
\Psi_\rhfeq(c) &= \det \begin{bmatrix}
A[c] & B[c] \\
C_{\bm{\theta}}(c) & F_{\bm{\theta}}(c)
\end{bmatrix}\\
\end{split}
\label{eqn:HFDS1}
\end{equation} 

In the case of HFDS (Eqn.~\eqref{eqn:HFDS1}), the NN parameters $\btheta$ and $A$ and $B$ are all optimized when
applied to the ground state approximation for Fermionic
Hamiltonian.

\section{Numerical Methods}{\label{sec:method}}
To support our statements about the relation among the NN wave functions, we will provide some numerical evidences. In this section, we will present the numerical methods for the ground state approximation of the Hubbard model, and the loss functions we define for mutual learning of wave functions, and the quantities for measuring the modulation on the variational wave function amplitudes.

\subsection{Model and Wave Function Architecture}
The variational wave functions are optimized to approximate the ground state of the Hubbard model
\begin{equation}
\hat{H}=-t\sum_{\langle i,j\rangle,\sigma}(\hat{c}_{i\sigma}^\dagger\hat{c}_{j\sigma}+h.c.)+U\sum_i \hat{n}_{i\uparrow}\hat{n}_{i\downarrow}  
\label{eq:hubbard}
\end{equation}
where $\hat{c}_{i\sigma}$ ($\hat{c}_{i\sigma}^\dagger$) is annihilation (creation) operator for spin-$\sigma$ ($\sigma=\uparrow,\downarrow$) electron on site $i$, and $\hat{n}_{i\sigma}=\hat{c}_{i\sigma}^\dagger \hat{c}_{i\sigma}$ is the number operator. We will consider this model on an $L=4\times 4$ square lattice with periodic boundary conditions on both spatial directions, interaction strength $U/t=8$, and $\frac18$ hole doping, i.e., $N_\uparrow=N_\downarrow=7$~\cite{ED_Hubbard}.

The neural network architecture is designed to have two hidden layers, each with 2,048 neurons and ReLU activation function, parametrized by $\btheta$, the set of weights and bias inside the NN. The output is an oversized array, which is reshaped into an $r\times N_s$ matrix $C_\btheta(c)$ and an $r\times r$ matrix $F_\btheta(c)$, which, together with the static matrices (with entries independent of $c$) $A$ and $B$, are used to construct the SPO and corresponding determinant matrices for various types of wave functions considered in this work (see Fig.~\pref{fig:keyPic}{e}).

\begin{figure}[h]
\centering
\begin{subfigure}[b]{0.48\textwidth}
\centering
\includegraphics[width=\textwidth]{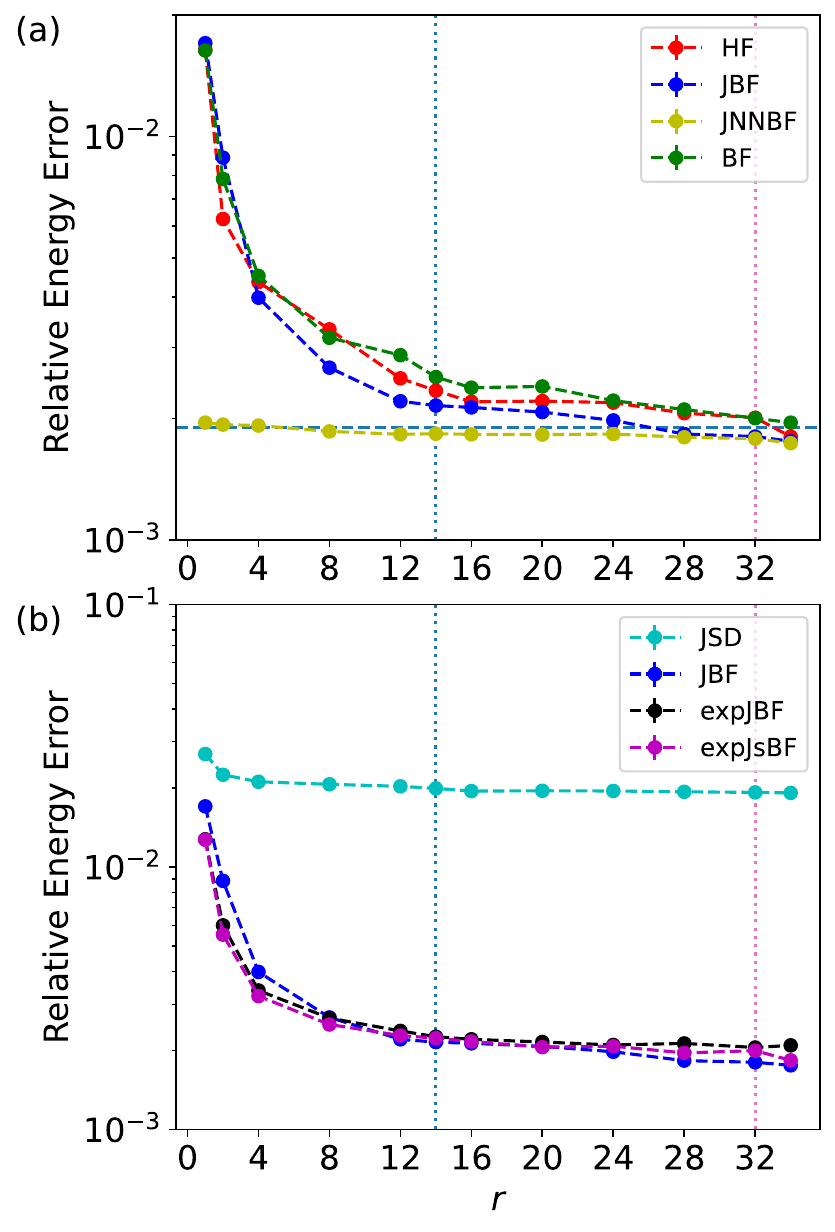}
\end{subfigure}
\caption{Variational energy results from ground state approximation on the Hubbard model ($L=4\times4$, PBC, $U/t=8$,  $N_{\uparrow}=N_{\downarrow}=7$). The two vertical dotted line show the locations of $r=N_s$(blue) and $r=L_s$(pink), and the horizontal dashed line in (a) is the lowest energy from $\rbfeq$ at $r=L_s$ from Fig.\ref{fig:fide-opt-wf-DC}. }
\label{fig:Energy-Rank}
\end{figure}

\subsection{Ground State Approximation}
To optimize the wave functions towards the ground state, in this work, we use supervised wave-function
optimization with Adam~\cite{kochkov2018variational,kingma2017adam} (SWO-Adam, see Appendix~\ref{append:swo}). Within SWO, at each so-called imaginary-time step $t$, we set a target state $\ket{\Phi_t}=e^{-\tau\hat{H}}\ket{\Psi_{t-1}}\approx (1-\tau\hat{H})\ket{\Psi_{t-1}}$ for the present variational wave function $\ket{\Psi_t}$. And a loss function to minimize in this procedure is defined as the logarithm of the wave-function fidelity:
\begin{equation}
\mathcal{L} = -\ln \frac{\langle\Psi_{t}\vert\Phi_t\rangle\langle\Phi_t\vert\Psi_{t}\rangle}{\langle\Psi_{t}\vert\Psi_{t}\rangle\langle\Phi_t\vert\Phi_t\rangle} 
\label{eq:LossFun}
\end{equation}
The gradient of Eqn.\eqref{eq:LossFun}
is estimated with Monte Carlo sampling from probability distribution $p(c)=|\Psi_{t-1}(c)|^2$,
\begin{equation}
\begin{split}    
\frac{\partial\mathcal{L}}{\partial\btheta}&\approx2\sum_{\{c|p(c)\}}\Re\left[\alpha(c)\Tr\left(\widetilde{A}_\btheta(c)^{-1} \frac{\partial\widetilde{A}_\btheta(c)}{\partial\btheta}\right)\right] 
\end{split}
\label{eq:GradEst}
\end{equation}
where the coefficient $\alpha(c)$ is given by
\begin{equation}
\alpha(c)=\frac{|\psi(c)|^2}{\sum_{\{c\}}|\psi(c)|^2}-\frac{\psi(c)^*\phi(c)}{\sum_{\{c\}}\psi(c)^*\phi(c)}   
\label{eq:alphacoe}
\end{equation}
\begin{equation}
\psi(c)=\frac{\Psi_t(c)}{\Psi_{t-1}(c)}, \quad \phi(c)=\frac{\Phi_t(c)}{\Psi_{t-1}(c)}   
\label{eq:ampratio}
\end{equation}
By iteratively feeding the estimated gradient Eqn.~\eqref{eq:GradEst} into Adam optimizer to minimize the loss function Eqn.~\eqref{eq:LossFun} at each step $t$, the variational state $\ket{\Psi_t}$ is undergoing imaginary time evolution stochastically, with energy being decreased until convergence.

The results, in terms of relative energy error to the exact ground energy for the chosen model $E_0=-11.868$~\cite{ED_Hubbard}, are shown in Fig.~\ref{fig:Energy-Rank}.

\subsection{Mutual Learning of Wave Functions}
The mutual learning of wave functions is the optimization of one wave function $\ket{\Psi_1}$, e.g., $\rjbfeq$ (see Sec.\ref{sec:NNBF_HF_SPO}), towards the other one $\ket{\Psi_2}$, e.g., $\rhfeq$. Two metrics are defined for this purpose: one is the wave function fidelity
\begin{equation}
\mathcal{F}_{\text{wf}}=\frac{\braket{\Psi_1}{\Psi_2}\braket{\Psi_2}{\Psi_1}}{\braket{\Psi_1}{\Psi_1}\braket{\Psi_2}{\Psi_2}}
\label{eq:fidewf}
\end{equation}
and the other is the square distance between the SPO's.
\begin{equation}
\mathcal{L}_{\text{spo}}=\frac1S\sum_{\{c\}}\left|\widetilde{A}_1(c)-\widetilde{A}_2(c)\right|^2    
\label{eq:spoloss}
\end{equation}
where $\widetilde{A}_{1/2}(c)$ are the $L_s\times N_s$-size SPO matrices generated from $\ket{\Psi_{1/2}}$ for configuration $c$, and $S$ is number of samples in $\{c\}$.  This latter metric is only applicable when both ansatz generate SPO's.  
We also defined the SPO fidelity as
\begin{equation}
\mathcal{F}_{\text{spo}}=\sum_{\{c\}}\frac{\det(\mathcal{S}_1(c)^{\dagger}\mathcal{S}_2(c))\det(\mathcal{S}_2(c)^{\dagger}\mathcal{S}_1(c))}{\det(\mathcal{S}_1(c)^{\dagger}\mathcal{S}_1(c))\det(\mathcal{S}_2(c)^{\dagger}\mathcal{S}_2(c))}    
\end{equation}
to measure the average overlap between the SPO's for each configuration $c$.

These metrics are used as the distances between two wave functions, and the target states are those from the ground state approximation for Hubbard model.

\subsection{Amplitude Modulation}
\label{sec:amp_mod}
To give an estimate for how rapidly various wave-functions oscillate both with respect to the magnitude and sign, we compute the following for different parts of the wave-function $f(c)$.

Given a Markov chain for Monte Carlo sampling from wave function $\ket{\Psi}$, we collect all the configuration samples from this chain removing successive identical configurations, such that the remaining ones in the set $\{c_j\}_{j=1}^J$  have the neighboring samples differing by the position of exactly one Fermion, i.e., $|c_j-c_{j+1}|=2$ for all $c_j$.  
We then consider  $\ln f_{\textrm{ratio}}(j) \equiv \ln f(c_{j+1}) - \ln f(c_j)$  with a particular focus on its average real and imaginary pieces, i.e.  
\begin{equation}
\mathcal{F}_{\text{magn}}=\frac{1}{J}\sum_{j=1}^J|\Re \ln f_\textrm{ratio}(j)|^2,\quad \mathcal{F}_{\text{phase}}=\frac{1}{J}\sum_{j=1}^J|\Im \ln f_\textrm{ratio}|^2    
\end{equation}
which are taken over the sample set $\{c_j\}_{j=1}^J$.   These quantities are independent of the norm of the wave-function and allow separate estimates of the magnitude and phase oscillations. 

\section{Comparison between Wave Functions}
\label{sec:NNBFvsHFDS}

We start by rewriting HFDS with $r$ hidden fermions in a somewhat different form.
Without loss of generality, we assume that $F_\btheta(c)$ is invertible for all $c$ (see Appendix~\ref{append:hfdsInvertF}).
Then we can rewrite  Eqn~\eqref{eqn:HFDS1} as 
\begin{equation}
\Psi_\rhfeq(c)=\det(F_\btheta(c))\det\left((A-BF_{\btheta}(c)^{-1}C_\btheta(c))[c]\right)   
\label{eq:rhf} \tag{$r$-HF}
\end{equation}

In the parlance of NNBF, the $\rhfeq$ wave-function is neural network backflow with an additive correction of the form 
\begin{equation}
\widetilde{A}_\btheta(c) = A-B(F_\btheta(c))^{-1}C_\btheta(c) \label{eq:rHFChange}
\end{equation}
multiplied by a determinant factor $F_\btheta(c)$  (alternatively an additive correction composed with a right-rotation with $F$).

This formula can be derived either directly by working with determinants of block matrices; or alternatively by using the matrix determinant lemma and computing the change in the determinant between the (padded with identity) matrix $A[c]$ and the matrix $(A+BC_\btheta(c))[c]$.

$\rhfeq$ then is a form of NNBF but does differ from the most standard Jastrow-NNBF in two ways.  
First, there is a somewhat non-standard ``Jastrow factor'' consisting of the determinant of a $r \times r$ matrix, which can be negative.  We will want to address two questions about this determinant factor:  (a) Is it preferable to use a determinant factor compared to a more basic (possibly also NN-inspired) Jastrow; (b) Does the addition of a determinant factor on top of NNBF span a larger class of wave-functions than no multiplicative factor at all.  
Second, we will be interested in understanding the difference in the additive corrections between $\rhfeq$ and standard NNBF.  Allowing the NNBF to use as its multiplicative factor an $r\times r$ determinant and fixing the NN's to the same depth and width, we want to ascertain whether using the NN output for matrices $C$ and $F$ which are multiplied as in Eqn.~\eqref{eq:rHFChange} to generate the additive SPO is better then just outputting $M$ directly.  

\subsection{Comparisons of Additive Corrections} 
{\label{sec:NNBF_HF_SPO}}

We will address these in reverse order starting first with comparing the additive correction between $\rhfeq$ and JNNBF.  

To make progress on this, it will also be useful to  define a closely related wave-function
\begin{equation}
\widetilde{A}_\btheta(c) = A + BC_\btheta(c) \tag{$r$-BF} \label{eq:BF}
\end{equation}
as well as its determinant-factor multiplied form $\rjbfeq$ (equivalently r-TBF). $\rjbfeq$ differs from $\rhfeq$ in that it directly uses the outputted matrix $C_\theta(c)$ whereas $\rhfeq$ uses  $(F_\theta(c)^{-1})C_\theta(c)$.

If we consider Eqn.~\eqref{eq:rHFChange} or Eqn.~\eqref{eq:BF}, we see that the additive correction to the SPO is the product of an $L_s \times r$ matrix ($B$) and an $r \times N_s$ matrix, ($F^{-1}_{\btheta}C_\btheta$ or $C_\btheta$), resulting in a low-rank additive correction of, at most, rank $r$.  As generically the rank of the additive correction to JNNBF will be full rank (i.e. $N_s$), almost none of the JNNBF SPO's will be representable by either Eqn.~\eqref{eq:rHFChange} or Eqn.~\eqref{eq:BF} when $r<N_s$.  On the other hand, we can show every additive SPO generated by $\rjbfeq$ can also be generated by JNNBF.  This is done by converting the $\rjbfeq$ neural network (even for $r>L_s$) into a JNNBF neural network of the same size by first increasing the $\rjbfeq$ neural network by one linear layer (encoding $B$ in the weights) so that it outputs $M_\btheta(c)$ and then compressing the two final linear layers into a single linear layer. This results in a JNNBF wave-function with the same initial layers. Notice, it is also straightforward to see that $L_s$-JBF and JNNBF span the same space (let $B$ be the identity) and that $(r+1)$-JBF contains $r$-JBF (but the converse is generically not true when $r<N_s$). In Fig.~\ref{fig:Energy-Rank}, we also see evidence of this numerically as the energy of $\rhfeq$ (or $\rjbfeq$) decrease monotonically with $r$ out to, at least, $r=L_s$ with the most significant decreases at $r<N_s$. See Appendix.~\ref{append:constructSubset} for more explicit arguments for these descriptions.

\begin{figure}[h]
\centering
\begin{subfigure}[b]{0.48\textwidth}
\centering
\includegraphics[width=\textwidth]{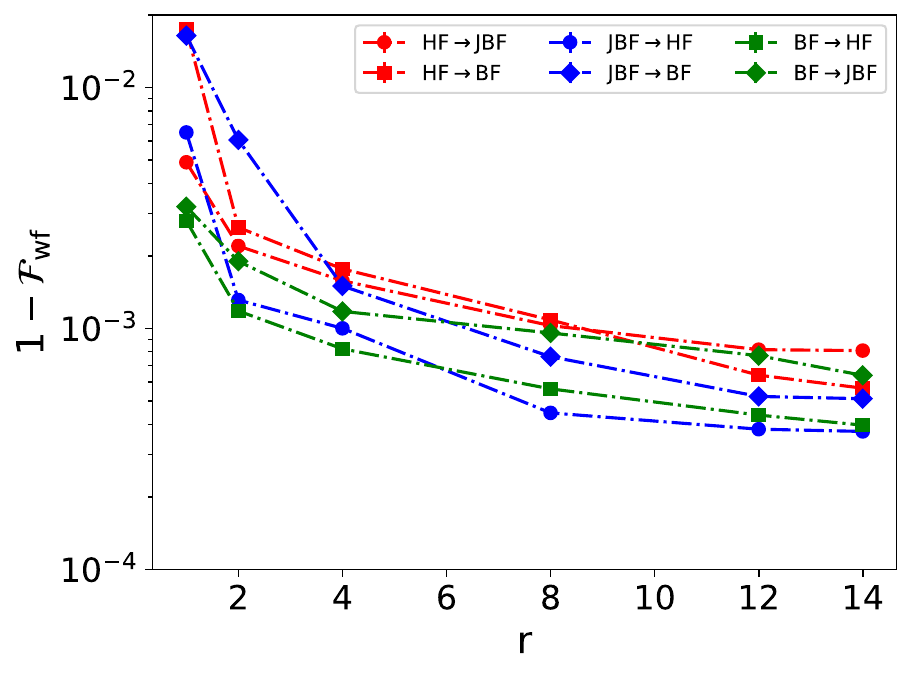}
\end{subfigure}
\caption{Wave function fidelity from mutual learning of different wave functions by \textit{maximizing the wave function fidelity}:  Optimization of \rhf towards \rjbf (red circle), and \rrbf (red square); \rjbf towards \rhf (blue circle), and \rrbf (blue diamond); \rrbf towards \rhf (green square), and \rjbf (green diamond). The target states are from the low energy space of Hubbard model ($L=4\times4$, PBC, $U/t=8$, $N_{\uparrow}=N_{\downarrow}=7$).}. 
\label{fig:fide-opt-wf}
\end{figure}

Analytically proving whether $\rhfeq$ is strictly included in JNNBF is made difficult by the $F^{-1}$ term. That said, JNNBF is almost certainly larger than $\rhfeq$ (for $r\leq L_s$) as we expect on general grounds that $\rjbfeq$ and $\rhfeq$ span a very similar (if not identical) space. In fact, in the $w \rightarrow \infty$ limit they do span the same set of configuration-dependent SPO's under the assumption of universal approximation theorem for arbitrary-width NN~\cite{pinkus_1999}.  At finite $w$, we have numerically  considered low-energy (with respect to the $4 \times 4$ Hubbard model) $\rjbfeq$ and $\rhfeq$ states. 
In Fig.~\ref{fig:fide-opt-wf}, we show the fidelity after mutual learning between $\rhfeq$ and $\rjbfeq$ ($r\leq N_s$) by maximizing the wave function fidelity Eqn.~\eqref{eq:fidewf}.
We found that $\rhfeq$ and $\rjbfeq$ can be optimized towards each other (in both directions) so that their infidelity is lower then $10^{-2}$.  This is consistent also with finding their optimized energies being very close with $\rjbfeq$ being slightly lower other than at $r=2$.
Unsurprisingly, we find that JNNBF finds lower-energy variational states than both $\rjbfeq$ and $\rhfeq$ at all $r$, as shown in Fig.~\ref{fig:Energy-Rank}. 
 
This leaves us with the following situation:
\begin{eqnarray}
\rhfeq & \approx   \rjbfeq\\
\rjbfeq & \underset{\text{spo}}\subset (r+1)\text{-JBF} \underset{\text{spo}}\subset L_s\text{-JBF}  \underset{\text{spo}}= \text{JNNBF} \label{eq:jbf_jnnbf_spo_relation} \\
r\text{-HF} & \underset{\text{spo}}\subset \text{JNNBF}:  w\rightarrow \infty 
\end{eqnarray}
where we use the notation $\underset{\text{spo}}\subset$ to indicate one class of wave-functions contains the same set of SPO's as the other class (a strictly stronger inclusion than the wave-functions being the same). Notice that Eqn.~\eqref{eq:jbf_jnnbf_spo_relation} is also true if we remove the determinant factor from each term.

\subsection{Benefit of Multiplicative Factor on Neural Network Backflow}
\label{sec:jastrow_det_benefit}
In the previous subsection, we argued that when dressed with a determinant factor, the additive corrections being used by JNNBF were generically a superset of the additive correction of other wave-functions such as $\rhfeq$ and $\rjbfeq$ whose additive corrections had restricted rank.  In this subsection, we will consider whether multiplying by a determinant out front - i.e. the determinant factor  - is important at all - i.e. is NNBF identical, in some limits, to JNNBF. 

To make progress on this, we will actually start by focusing on comparing $\rjbfeq$ (or $\rhfeq$) with $\rbfeq$. To make this comparison, we will take the $\rjbfeq$ ($\rhfeq$) wave-functions and rewrite them as a BF wave-function with a different additive correction. -i.e. we will have the multiplicative factor absorbed into the additive correction. 

When $r\leq N_s$, we can rewrite  (see Appendix~\ref{append:JastrowAbsorbSPO} for details) these wave-function as 
\begin{equation}
\Psi_{r\text{-HF}/\text{JBF}}=\det\left(  (A-B'C_{\bm{\theta}}'(c))[c] \right)   
\label{eq:rjbf_to_2rbf}
\end{equation}
where 
\begin{equation}
B' = 
\begin{bmatrix}
-A_1 & B    
\end{bmatrix},
\label{eq:rjbf_to_2rbf_b}
\end{equation}
and 
\begin{equation}
C_\btheta'(c) = 
\begin{cases}
\begin{bmatrix}
F-I & \bm{0} \\
F^{-1}C_1F & F^{-1}C_2
\end{bmatrix}, & \text{for } r\text{-HF}\\
\begin{bmatrix}
F-I & \bm{0} \\
C_1F & C_2
\end{bmatrix}, & \text{for } r\text{-JBF}
\end{cases}
\label{eq:rjbf_to_2rbf_c}
\end{equation}
are of size $L_s\times 2r$ and $2r\times N_s$, respectively.
Notice that, this means each $\rjbfeq$ ($\rhfeq$) wave-function can be written as a slightly more complicated additive correction of rank $2r$.   
As both $2r$-BF and $\rjbfeq$ can be written as a neural network backflow with rank-$2r$ additive corrections, this suggests that they might span a similar space of wave-functions. Generically, we might expect that $2r$-BF actually spans a larger space because the $2r \times L_s$ SPO's are completely general for $2r$-BF while for $\rjbfeq$ ($\rhfeq$) the upper right $r \times (N_s-r)$ corner of the additive correction is forced to be zero.  In the $w \rightarrow \infty$ limit it is strictly true that $2r$-BF contains the (determinant absorbed) SPO's of both $\rjbfeq$ and $\rhfeq$ (but not the converse).  

\begin{figure}[h]
\centering
\begin{subfigure}[b]{0.48\textwidth}
\centering
\includegraphics[width=\textwidth]{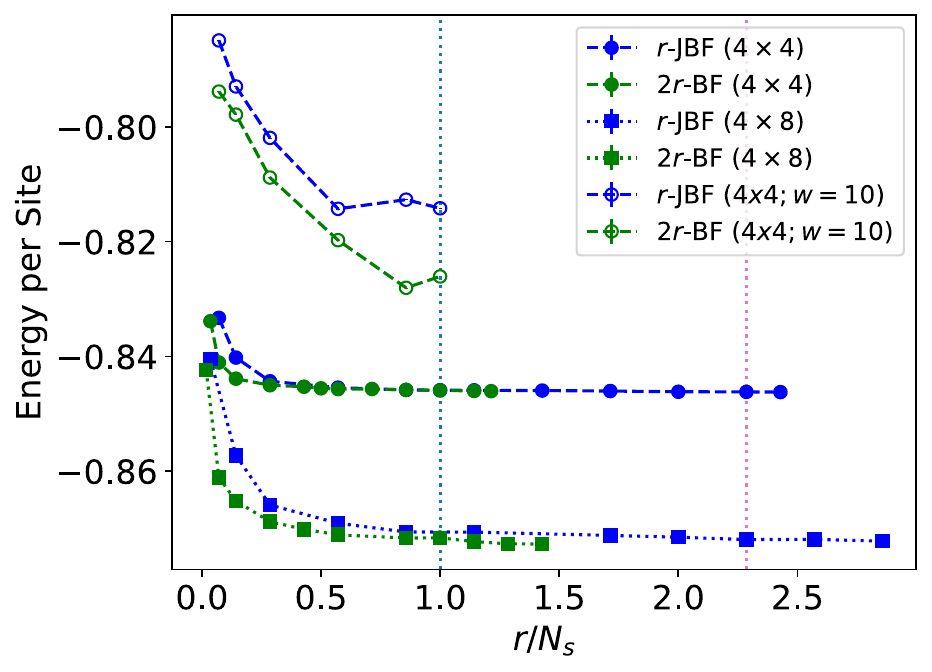}
\end{subfigure}
\caption{Replot of the energy data from Fig.~\ref{fig:Energy-Rank} for comparison between $\rjbfeq$ (blue) and $2r$-BF (green). In addition to the results for previous model ($L=4\times4$, PBC, $U/t=8$,  $N_{\uparrow}=N_{\downarrow}=7$)(circle markers), the results from a larger lattice ($L=4\times8$, PBC, $U/t=8$,  $N_{\uparrow}=N_{\downarrow}=14$)(square markers), and results for small-size NN (width=10)(empty circle markers) are displayed as well.}
\label{fig:EnergyBD-Rank}
\end{figure}

Unsurprisingly, we find that energetics of $2r$-BF and $\rjbfeq$ are similar on both $4 \times 8$ and $4 \times 4$ Hubbard models (see Fig.~\ref{fig:EnergyBD-Rank}) with $2r$-BF being slightly lower in energy. 

\begin{figure*}
\centering
\includegraphics[width=\textwidth]{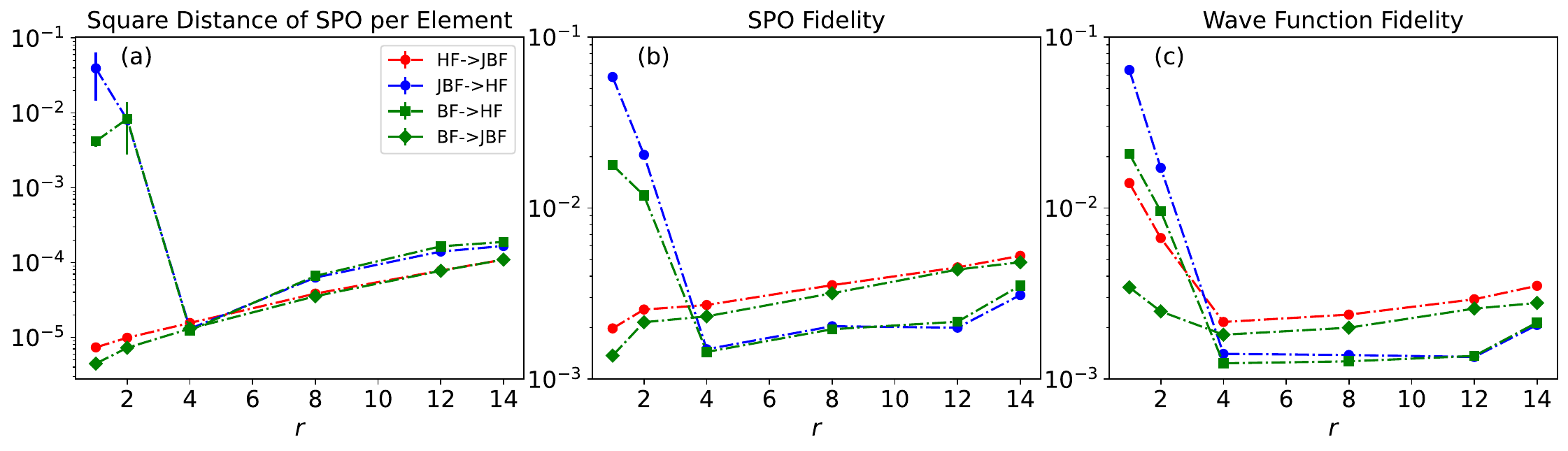}
\caption{(a) Square distance of SPO's per configuration and matrix element, (b) SPO fidelity ($1-\mathcal{F}_{\text{spo}}$ in the plot) and (c) wave function fidelity ($1-\mathcal{F}_{\text{wf}}$ in the plot) from mutual learning of different wave functions by \textit{minimizing the SPO distance}: Optimization of \rhf towards \rjbf (red circle); \rjbf towards \rhf (blue circle); \rrbf towards \rhf (green square), and \rjbf (green diamond). The target states are from the ground state approximation for Hubbard model ($L=4\times4$, PBC, $U/t=8$, $N_{\uparrow}=N_{\downarrow}=7$).}
\label{fig:spo-fide}
\end{figure*}

Moreover, through numerical fidelity matching of the low energy states (see Fig.~\ref{fig:fide-opt-wf}), we find that $2r$-BF can represent both $\rjbfeq$ and $\rhfeq$ with infidelities less then $10^{-2}$ for all $r$.  In the case of $\rjbfeq$, essentially this tells us that $2r$-BF can output the matrix $C_1F$ from $\rjbfeq$ where previously it was only outputting $F$ and $C$ separately. The converse ($\rjbfeq$ and $\rhfeq$ matching $2r$-BF) also achieves infidelities less then $10^{-2}$ except at $r=1$ where the fidelity matching is worse in that direction.   

We even attempt to go beyond just fidelity matching and try to directly optimize the SPO distance between $2r$-BF and $\rjbfeq$ by optimizing the $2r$-BF wave-function (the opposite is not possible because of the zeros). 
In Fig.~\ref{fig:spo-fide}, we also show the results from mutual learning by minimizing the SPO distance Eqn.~\eqref{eq:spoloss}. Notice that optimizing SPO distance is more strict than optimizing the wave function fidelity, as there is a significant amount of degree of freedom for the SPO to give the same amplitude during the operations of the row selection and determinant evaluation. Hence, in Fig.~\pref{fig:spo-fide}{b,c} we further show the SPO fidelity and the wave function fidelity that are consistent with the SPO distance in Fig.~\pref{fig:spo-fide}{a}.
In some limit, optimizing the SPO distance and the fidelity are closely related (i.e. their respective infidelities become zero at the same time), but they do probe the closeness of wave-functions in different ways. For example, it is possible to have low SPO-distance but not super-high fidelity because even if the SPO distance is small, its contribution to the fidelity can be amplified by large values generated from the inverse of the row-selected SPO (see Appendix.~\ref{append:fide_spo_relation} for a detailed discussion of this). We find that the $2r$-BF can match the SPO's quite successfully at all $r$'s for $\rjbfeq$ and at $r \geq 4$ for $\rhfeq$. 

While the above numerical experiments are performed by fixing the NN for all the wave functions, we can also ask whether there are `minor' architectural changes to the FFNN for $2r$-BF which would allow for an exact reconstruction of other wave-functions.
We demonstrate an example of this reconstruction in Appendix.~\ref{append:rjbf_from_2r_bf} where we write a $\rjbfeq$ wave-function as a $\rbfeq$  wave-function  with two additional layers (with different activation functions) and widths expanded only by an extra factor of $r$.  

It is worth mentioning that it is also possible to reconstruct $\rhfeq$ from $2r$-BF by adding a few more layers to the FFNN, where we can use adjugate formula to obtain the inverse of $F$, i.e., $F^{-1}=\text{adj}(F)/\det F$. However, this involves the evaluation of determinants within the NN, and it will require widths on the order of $r!$ number of neurons if implemented directly via the definition of determinant making this direct approach impractical (we know that the evaluation of determinant can be done in $O(n^3)$ steps, but this will require many more layers within the NN).   Nevertheless, the fidelity and the energetics from Fig.~\ref{fig:Energy-Rank} indicate that even without any change in architecture, $\rjbfeq$ or NNBF are already as expressive as $\rhfeq$ and effective in approximating the ground state without the necessity of $F^{-1}$ to be included in the additive correction.  

\begin{figure}[h]
\centering
\begin{subfigure}[b]{0.48\textwidth}
\centering
\includegraphics[width=\textwidth]{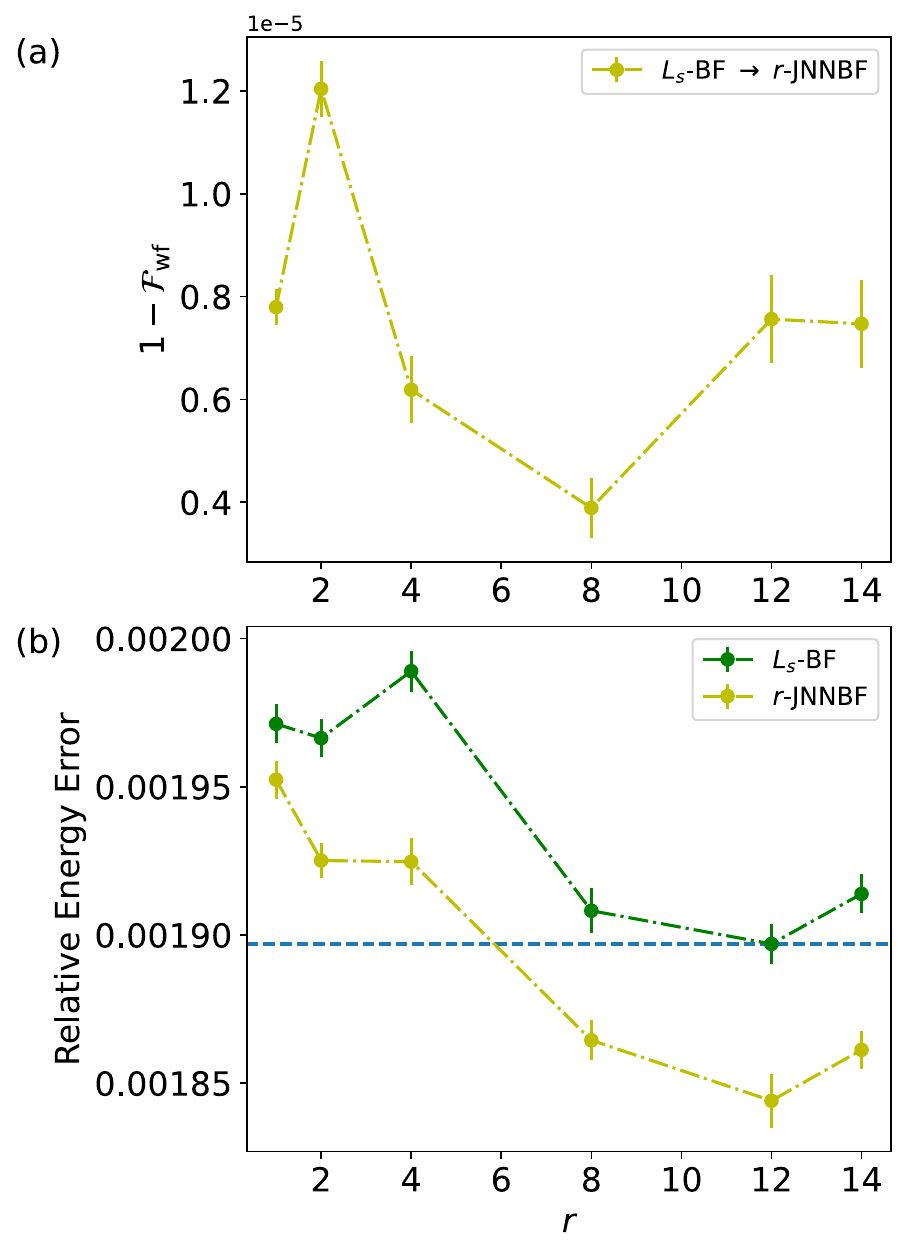}
\end{subfigure}
\caption{Results of mutual learning of $L_s$-BF towards $r$-JNNBF by \textit{maximizing the wave function fidelity}. (a) Wave function fidelity (b) variational energy after the optimization. The horizontal lines in (b) is the lowest energy of $L_s$-BF after the optimization. }. 
\label{fig:fide-opt-wf-DC}
\end{figure}

We now directly consider the relation between NNBF and $\rjbfeq/\rhfeq$. On the one hand, when $2r<L_s$, almost none of the NNBF SPO's will be representable by either of these two wave-functions. On the other hand, we already know that all the $2r$-BF wave-functions are representable by NNBF and given the evidence that $2r$-BF is a superset of $\rjbfeq$, this suggests that all ($r\leq N_s)$ $\rjbfeq$ have (determinant absorbed) SPO's that are representable by NNBF; this is rigorously true for $w\rightarrow \infty$.  This argument also suggests that JNNBF and NNBF should span the same space for all $r\leq N_s$.  Upon optimization (see Fig.~\ref{fig:Energy-Rank}), we find that their energetics are extremely close with very minimal $r$ dependence in JNNBF out to $r=N_s$ (surprisingly actually even for $r>N_s$ there is still very little $r$ dependence). We again attempt to match the fidelity of $r$-JNNBF with NNBF; This optimization is done through $L_s$-BF which is equivalent to NNBF. In Fig.~\ref{fig:fide-opt-wf-DC}, we show the fidelity after learning $\rjnnbfeq$ ($r\leq N_s$) with $L_s$-BF, and the comparison between energy after the optimization. And we find an infidelity of essentially $10^{-5}$ for all $r$, suggesting that NNBF really does contain JNNBF. 

The conclusion of this section is that including a determinant factor which multiplies an NNBF wave-function seems to not increase the space of wave-functions spanned by NNBF , and increases the space of wave-functions spanned by $\rbfeq$ up to $2r$-BF when $r\leq N_s$.  When the number of hidden fermions is very large - more then the number of actual fermions, it is less clear whether the determinant factor should have additional benefit although in practice we see almost no decrease in energy for JNNBF at $r>N_s$ (See Fig.~\ref{fig:Energy-Rank}).

\subsection{Exponential Jastrow vs Determinant Factor}
What we have seen in the previous section, is that the determinant factor matters very little for JNNBF, but does increase the `rank' or accessible matrix size of BF from $r$ to $2r$.  Also, when $r>N_s$, formally the determinant factor can be important.  
If one is going to use a multiplicative factor to enhance the wave-function, it's worth considering whether the determinant factor is better than a more standard alternative of a Jastrow factor.  In particular, a reasonable comparison is to use an exponential Jastrow where the term being exponentiated also comes from NN;  we call this class of wave-functions expJBF. Like a standard Jastrow, we have that there is always a positive correction to the determinant backflow.  We can additionally test whether this is important by one more multiplicative factor, expJsBF which multiplies the Jastrow factor by an additional NN-generated scalar that is allowed to be positive and negative. 

In this section, we primarily compare the energetic of these wave-functions, to numerically check which have lower energy on our prototypical $4 \times 4$ Hubbard model (See Fig.~\pref{fig:Energy-Rank}{b}). We find that the energy of expJBF and expJsBF are almost identical although the signed version gives slightly better results at large $r$. In comparison against JBF, we find that at small rank, we get lower energy from the exponential Jastrows than the determinant factor; this relationship seems to switch at large enough $r$.  

A plausible mechanism for this difference is that the role of the Jastrow or the multiplicative factor is to be able to introduce exponentially large separation between different configurations. An exponential Jastrow naturally has this ability;  the determinant factor essentially emulates this exponential separation through the product of the respective matrices eigenvalues.  At low $r$, it is restricted in the number of such terms in the product and has trouble matching the exponential Jastrow. This all said, the difference between the various multiplicative factors are tiny especially when $r$ is large, and it is not clear that one should take those tiny differences at this level seriously. The high-level result is that some multiplicative factor marginally improves a restricted form of the NNBF wave-function, but different reasonable multiplicative factors lead to essentially the same marginal improvement;  there seems to be nothing fundamentally special or powerful about the determinant factor.  

\subsection{Determinant Factor vs Effective Determinant Factor}
Throughout this paper, we've been making a somewhat artificial distinction between the determinant factor and Slater Determinant pieces of the wave-functions.  There is a sense in which one could make a different decomposition of the wave-functions we've been considering, by writing the wave-function as a fixed Slater determinant by an additional "effective determinant factor".
In particular, we can write all the wave-functions considered so far as 
\begin{equation}
\Psi(c) = \det(A[c])\det(\bar{F}(c)) 
\end{equation}
where $\bar{F}$ absorbs the rest of the wave-function except the static Slater determinant part $A[c]$. Notice that (1) $\bar{F}$ will depend on the NN in some way and (2) unlike all the other multiplicative factors we've considered, the matrix $\bar{F}$ for which we are taking the determinant of may involve row-selection.  
In particular, we can write NNBF as
\begin{equation}
\Psi_{\text{NNBF}}(c)=\det(A[c])\det(I-CA[c]^{-1}B[c])   
\end{equation}
and r-HF as 
\begin{equation}
\Psi_{\rhfeq}(c)= \det(A[c])\det(F-CA[c]^{-1}B[c])
\end{equation}

Naively, it's not obvious why (or if) these wave-functions should be superior to a Slater Determinant dressed with a determinant factor - i.e. $\bar{F}=F_\btheta(c)$ which we call JSD. Numerically, we have examined this wave-function and find JSD is much worse in energy than any of the NNBF-type wave-functions.  It is naively surprising that the NN in JSD is unable to learn $F_\theta(c) = \det(I-CA[c]^{-1}B[c])$ at our given width; 
 notice  in the $w \rightarrow \infty$ limit JSD could learn that $F_\btheta(c)$. 

To understand the mechanism for this distinction, let us first look at the determinant factor $\det F(c)$.  The NN, as well as the determinant function, is a smooth function where small change of the input value will only cause a small change to the output; the determinant factor will not change drastically as the input $c$ moves between the nearest configurations. However, the amplitude of the exact ground state is not necessarily a smooth function of the configurations, thus is likely to go beyond the modulation ability of a determinant factor at moderate $w$. 

On the other hand,  the effective determinant factor from NNBF $\det(I-CA[c]^{-1}B[c])$, has the presence of $A[c]^{-1}B[c]$, which is essentially not a smooth function of input configurations due to row-selection and will impose a drastic change on top of the smooth change from the NN, making it possible to capture the pattern from the true ground state, while still maintaining the generalization ability of the NN even at small $w$. 

We can explicitly test and quantify whether the effective determinant factor with row-selection generates significantly stronger modulation than directly using the NN output as a determinant factor. To accomplish this, we want to look at the difference in amplitudes for configurations that are "nearest neighbors" (see Sec.~\ref{sec:amp_mod} for details).  

Here the following quantities will be considered: (1) $f(c)=\det F_{\btheta}(c)$, where $\det F_{\btheta}(c)$ is the determinant factor from the definition of wave function ansatz, e.g., the one from Eqn.~\eqref{eq:rhf}; (2) $f(c)=\det \bar{F}(c)$, the effective multiplicative factor obtained by factoring out the static Slater determinant part of the wave function, e.g., $\det \bar{F}(c)=\det \left(F_\btheta(c)-C_\btheta(c)A[c]^{-1}B[c]\right)$ for $\rhfeq$; (3) $f(c)=\mathcal{S}_{kj}(c)$, the matrix element from the SPO. 

First, we look at a prototypical example for $\bar{F}$ of both $\rhfeq$ and $r$-JSD at $r=N_s$ where we measure the ratio of nearest-neighbor amplitudes of the effective determinant factor, $f_\text{ratio}(c_j)=\det\bar{F}(c_{j+1})/\det\bar{F}(c_j)$ for random configurations sampled from $|\Psi(c)|^2$.  

\begin{figure*}
\centering
\includegraphics[width=\textwidth]{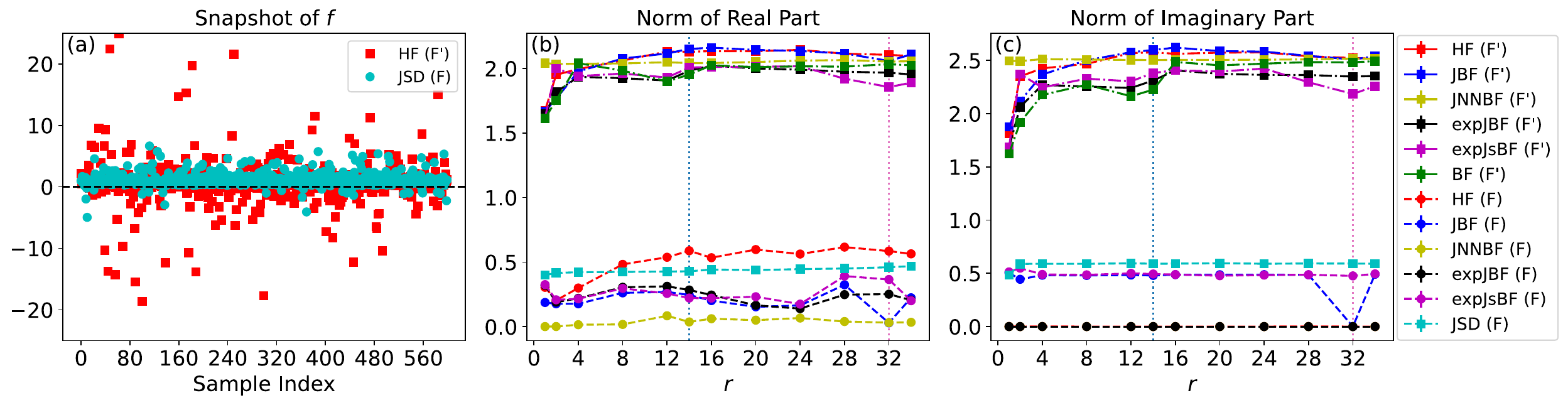}
\caption{(a) Samples of $f_{\text{ratio}}$ for the effective determinnat factor from $\rhfeq$ and $r$-JSD ($r=N_s$) along a Markov chain. (b) The average square norm of the real part of $f'$. (c) The average square norm of the imaginary part of $f'$. In (b) and (c), the results for the defined multiplicative factors ($F$) and the effective multiplicative factors ($F'$) are plotted for all the wave functions considered in this paper. The wave functions are from the ground state approximation for Hubbard model ($L=4\times4$, PBC, $U/t=8$, $N_{\uparrow}=N_{\downarrow}=7$).}
\label{fig:unif_fide}
\end{figure*}

In Fig.~\pref{fig:unif_fide}{a}, we can readily see an obvious distinction between the $f_\text{ratio}$, from $\rhfeq$ and $\rjsdeq$. More specifically, the ratios from $\rjsdeq$ show that the determinant factor, $\det\bar{F}=\det F_\btheta$, modulates the  wave function amplitude in a weak way for both the sign structure and the magnitude. In contrast, the effective determinant factor from $\rhfeq$ modulates both the magnitude and the sign structure in a strong way. 

We can further quantify this modulation by computing the real and imaginary components of the log-ratio $\ln f_\text{ratio}(c_j)$ which act as an effective derivative, and the norms of them measure the modulation on the magnitude and sign structure respectively. Using the log importantly lets us see rapid relative changes in the amplitude even when $|f_\text{ratio}|\ll1$.  
In Fig.~\pref{fig:unif_fide}{b,c}, we again observe a sharp distinction between the effective determinant factors and determinant factors from all types of the wave functions considered in this work. Note that the larger the norm is, the stronger the modulation is, so that the modulation from effective determinant factor is nearly $\exp(0.75)\approx 2.1$ times stronger, based on the differences of the average norms in Fig.~\pref{fig:unif_fide}{b,c}, than the determinant factor on both magnitude and sign structure. Interestingly, we can also observe from Fig.~\pref{fig:unif_fide}{b,c} that at lower rank, the effective determinant factor modulations from $\rhfeq$, $\rjbfeq$ and $\rbfeq$ are weaker than their counterparts at higher rank, which is consistent with their relatively higher energies in  Fig.~\ref{fig:Energy-Rank}.   

\begin{figure*}
\centering
\includegraphics[width=\textwidth]{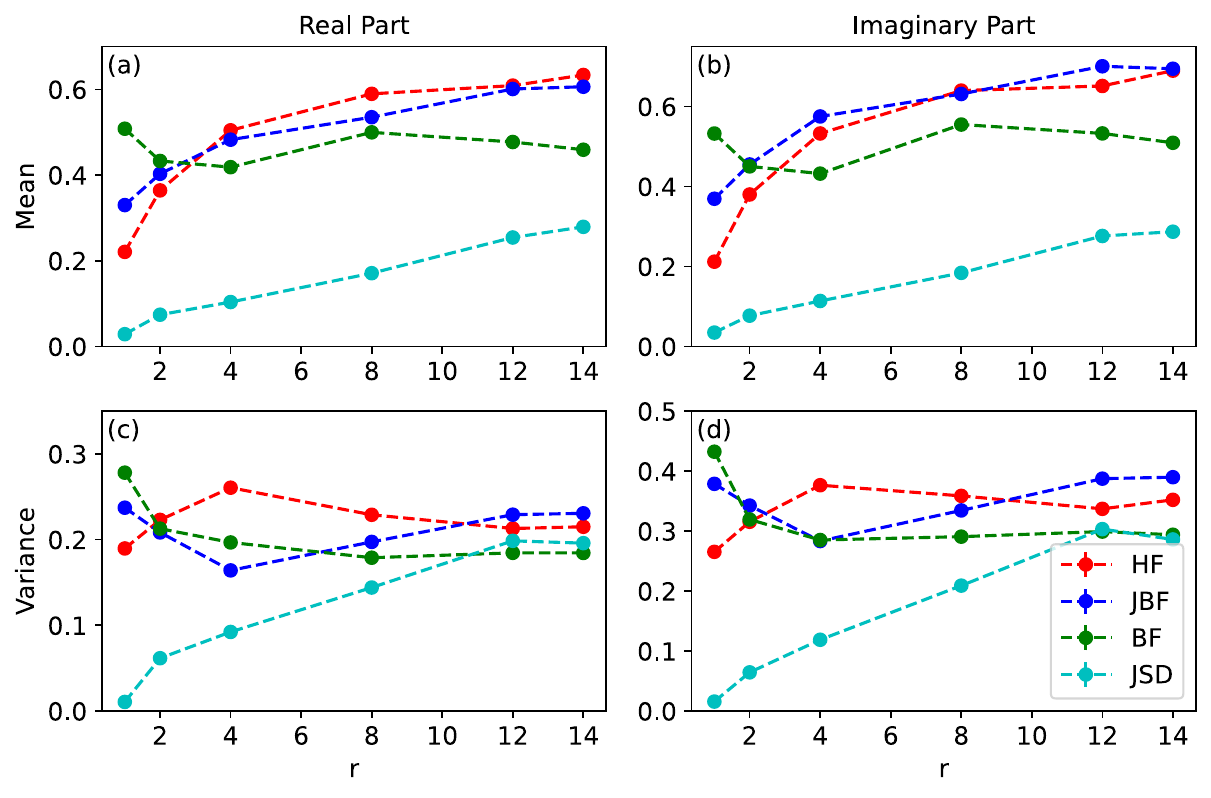}
\caption{The average norm of effective derivative on the SPO elements for $\rhfeq$, $\rjbfeq$, $2r$-BF and $r$-JSD. (a) The mean of $\mathcal{F}_{\text{magn}}$ over all the SPO elements; (b) The mean of $\mathcal{F}_{\text{phase}}$; (c) The variance of $\mathcal{F}_{\text{magn}}$ over all the SPO elements; (d) The variance of $\mathcal{F}_{\text{phase}}$; The wave functions are from the ground state approximation for Hubbard model ($L=4\times4$, PBC, $U/t=8$, $N_{\uparrow}=N_{\downarrow}=7$).}
\label{fig:unif_spo}
\end{figure*}

Alternatively, we compute the log-ratio of the matrix elements from SPO, and show the mean and variance over these $L_s\times N_s$ elements in Fig.~\ref{fig:unif_spo}. From them, we can extract some information about the modulation on the SPO from various types of wave function ansatz. Although the construction of SPO does not involve the row-selection as the wave function amplitude, we still observe weaker modulations from JSD. The mean for JSD is always lower than those from NNBF-type wave functions, in both magnitude (real part of log-ratio) and sign (imaginary part of log-ratio) modulation, indicating that on each SPO element, JSD changes the values in a relatively weaker way. Moreover, the smaller variance from JSD shows that JSD impose a more uniform modulation over the entire SPO, suggesting that JSD is less capable of identifying which part of SPO is worth stronger modulation than the rest.    

Considering the geometry for the SPO, it is easy to see that, in the cases of $r\leq N_s$, the SPO of JSD is always within the linear space spanned by the column vectors from the static matrix $A$, while NNBF is dynamically changing the space for the SPO given different configuration inputs. In Appendix.~\ref{append:spo_jsd_nnbf}, we demonstrate that it is impossible for JSD to reproduce the SPO from NNBF.  

Overall we see that the effective determinant factor with row-selection imposes stronger modulation on the wave-function amplitudes, achieving better performances on ground state approximation. The investigation on SPO also suggests that NNBF produces a broader class of SPO, that goes beyond the representability of Slater-Jastrow type wave functions, indicating that NNBF is likely to be a better ansatz candidate for Fermionic systems. 

\section{Summary and Overview}
In this work, we have considered a series of different NNBF wave-functions and their relation with HFDS.  We show that HFDS can be written in the NNBF form with an $r\times r$ determinant factor, $\det F_\btheta(c)$,  and an additive correction to the SPO of the form  $BF_\btheta(c)^{-1}C_\btheta(c)$, where $F^{-1}C$ is an $r\times N_s$ matrix (see Fig.~\pref{fig:keyPic}{a} bottom and Eqn.~\eqref{eq:rhf}).  This is not of full-rank for $r\leq N_s$ and is to be contrasted with the standard NNBF where the full-rank additive correction (an $L_s \times N_s$ matrix) is generated directly.

The determinant factor piece turns out to not be critical for NNBF wave-functions: it can be replaced by an exponential Jastrow which achieves similar (and often better) energetics (see Fig.~\pref{fig:Energy-Rank}{b}) or alternatively when $r\leq N_s$ it can be absorbed into the additive correction to the SPO resulting in a new more complicated additive correction $BC’$ where $C’$  is now “taller” by size $r$  (see r.h.s of Fig.~\pref{fig:keyPic}{a,b}) and Eqns.~\eqref{eq:rjbf_to_2rbf}-\eqref{eq:rjbf_to_2rbf_c}).

We are then left comparing various different additive corrections of the form $BC’$  where $C’$ is generated from various combinations of neural network output.  Two general principles appear here.  First and unsurprisingly, larger $r$ is better until $r=L_s$.  Larger $r$ allows access to more flexibility in the matrices that can be generated including, out to $r=N_s$, access to higher rank matrices. Beyond $r\geq L_s$, we can reduce additive corrections to $r=L_s$.  Secondly, the different ways we’ve considered to use the NN output to produce a $C’$ of size $2r$ (i.e. $r$-JBF, $r$-HF, $2r$-BF, etc) all are roughly similar in energetics and expressibility and, where they differ, favor the states which are simpler, more direct, and more generic updates (i.e. $2r$-BF). In particular, we see that $r$-HF and $r$-JBF are empirically very close energetically (with a slight edge to $r$-JBF, see Fig.~\pref{fig:Energy-Rank}{a}) and that we can optimize them towards each other with relatively low infidelity in each direction(see Fig.~\ref{fig:fide-opt-wf}).   $2r$-BF is consistently lower in energy than both $r$-JBF and $r$-HF(see Fig.~\ref{fig:EnergyBD-Rank}) and optimizing toward and from $2r$-BF can be done with low infidelity (see Fig.~\ref{fig:fide-opt-wf}).  In the limit of large width, $2r$-BF can represent any SPO in $r$-JBF and $r$-HF; and with two extra layers can represent any SPO in $r$-JBF at finite width. Amongst the various NNBF wave-functions, these two principals suggest that using standard NNBF is better energetically and likely spans the space of all the other wave-functions we’ve considered when $r\leq N_s$.  For $r>N_s$,  it likely makes sense to use NNBF with either an exponential Jastrow or determinant factor (although the improvement in energy even then is somewhat marginal).

NNBF and HFDS have both been successful in capturing highly accurate ground states.  We compare a wave-function JSD which is similar on the surface, a determinant factor times a Slater Determinant, and find that it is significantly worse.  To help understand what aspects are endowing NNBF with its particular advantage, we rewrite all our wave-function as a (non-configuration independent) Slater Determinant times an “effective determinant factor” and then compare properties of these effective factors.  In cases, such as NNBF and HFDS where the effective determinant factor involves row-selection, we see that both the sign and amplitude are rapidly oscillating as a function of configuration, something which does not happen at moderate width for JSD. This suggests that some of the advantage of NNBF is coming from the ability to rapidly change the wave-function.

Unlike the tensor network ansatz, which are well understood in terms of entanglement~\cite{White1992,Verstraete2006,Vidal2007,tensornet2019,AreaLaw2009,ORUS2014117,mera2008}, the development of neural quantum states (NQS) largely relies on heuristic experiments, and lacks a comprehensive understanding for the underlying principles~\cite{Deng2017,Passetti2023,denis2023comment,trigueros2023meanfield,Sharir2022,Clark2018,abrahamsen2023anti}. We hope our work paves the way for unifying various types of NQS and identifying the crucial components for an efficient wave-function ansatz, which, in turn, may lead to the discovery of more accurate and scalable NQS.

\appendix

\begin{acknowledgments}
We acknowledge the helpful conversations with A. Liu, D. Luo and K. Loehr. 
This work made use of the Illinois Campus Cluster, a computing resource that is operated by the Illinois Campus Cluster Program (ICCP) in conjunction with the National Center for Supercomputing Applications (NCSA) and which is supported by funds from the University of Illinois at Urbana-Champaign. We acknowledge support from the NSF Quantum Leap Challenge Institute for Hybrid Quantum Architectures and Networks (NSF Award 2016136).
\end{acknowledgments}
\section{Validity of Eqn.~\eqref{eq:rhf}}
{\label{append:hfdsInvertF}}
In Sec.~\ref{sec:NNBFvsHFDS}, we write down the transformed wave function expression for HFDS, Eqn.~\eqref{eq:rhf}, by assuming that the backflow matrix $F_\btheta(c)$ is invertible. Here we are proving that even if originally $F_\btheta(c)$ is not invertible for all the configuration input $c$'s, we can always construct an exact transformation (without changing the amplitude output or the neural network architecture) on HFDS, such that 
\begin{enumerate}
\item $F_\btheta(c)$ is invertible for all $c$'s when the backflow matrix $\begin{bmatrix}
C_\btheta(c) & F_\btheta(c)    
\end{bmatrix}$ for $\rhfeq$ is at least rank $r$.
\item Eqn.~\eqref{eq:rhf} is still valid for all $c$'s when there exists backflow matrix $\begin{bmatrix}
C_\btheta(c) & F_\btheta(c)    
\end{bmatrix}$ for $\rhfeq$ with rank less than $r$.  
\end{enumerate}

Given the original HFDS wave function form
\begin{equation}
\begin{split}
\Psi_\rhfeq(c) &= \det \begin{bmatrix}
A[c] & B[c] \\
C_\btheta(c) & F_\btheta(c)
\end{bmatrix}
\end{split}
\label{eqn:hfds_adx1}
\end{equation} 
We can divide the configuration basis sets into two disjoint sets: invertible set $V$ whose elements $c$ give invertible $F_\btheta(c)$, and uninvertible set $\Bar{V}$ with uninvertible $F_\btheta(c)$, or equivalently, $\det F_\btheta(c)=0$. The underlying reason for $\det F_\btheta(c)=0$ is that the $r\times r$ matrix $F_\btheta(c)$ is made up of less than $r$, say, $r_c$, linearly independent vectors. Performing Gaussian elimination on the columns of $F_\btheta(c)$, we will end up with
\begin{equation}
F_\btheta(c)\rightarrow \begin{bmatrix}
\widetilde{f}_{1,\btheta}(c) &\widetilde{f}_{2,\btheta}(c) & \cdots & \widetilde{f}_{r_c,\btheta}(c) & 0 & \cdots & 0 
\end{bmatrix}
\label{eqn:f_gausselimi}
\end{equation}
where $\{\widetilde{f}_{j,\btheta}(c)\}_{j=1}^{r_c}$ are $r$-dimensional column vectors. Because of the presence of zero vectors in Eqn.~\eqref{eqn:f_gausselimi}, we get the unwanted $\det F_\btheta(c)=0$ and uninvertible $F_\btheta(c)$. Nevertheless, suppose that a $c\in \Bar{V}$ has backflow matrix $\begin{bmatrix}
C_\btheta(c) & F_\btheta(c)    
\end{bmatrix}$ is of rank at least $r$, but unfortunately has rank-deficient $F_\btheta(c)$ as given by Eqn.~\eqref{eqn:f_gausselimi}, we can mix it with new linearly independent vectors from $C_\btheta(c)$ to remove those zero vectors, such that $F_\btheta(c)$ is full-rank and invertible. To do so, we introduce an $(N_s+r)\times(N_s+r)$ mixing matrix $K$, which is upper-triangular with 1's on the diagonal:
\begin{equation}
K=
\begin{bmatrix}
1 & k_{12} & \cdots & k_{1,N_s+r} \\
0 & 1 & \cdots & k_{2,N_s+r} \\
0 & 0 & \ddots & \vdots \\
0 & 0 & \cdots &  1
\end{bmatrix}    
\end{equation}
We right multiply the $\rhfeq$ matrix with this mixing matrix $K$, the result of which will be transparent if we rewrite the $\rhfeq$ matrix in terms of column vectors:
\begin{equation}
\begin{split}
&\begin{bmatrix}
A[c] & B[c] \\
C_\btheta(c) & F_\btheta(c)
\end{bmatrix}K \\
&=
\begin{bmatrix}
h_1 & \cdots & h_{N_s} & f_1 & \cdots & f_r    
\end{bmatrix}K \\
&=
\begin{bmatrix}
\overline{h}_1 & \cdots & \overline{h}_{N_s} & \overline{f}_1 & \cdots & \overline{f}_r
\end{bmatrix}
\end{split}    
\end{equation}
where
\begin{equation}
\overline{h}_{m}=\sum_{j=1}^{m-1}k_{j,m}h_j+h_m 
\end{equation}
\begin{equation}
\overline{f}_{m}=\sum_{j=1}^{N_s}k_{j,N_s+m}h_m+\sum_{j=1}^{m-1}k_{N_s+j,N_s+m}f_{j}+f_m     
\end{equation}
Effectively, the $K$ matrix adds to each column vector a linear combination of previous ones. By choosing appropriate matrix entries for $K$, the right multiplication of $K$ would increase the rank of $F_\btheta(c)$ up to $r$ (full-rank) via adding new column vectors from $C_\btheta(c)$. As long as those $c\in\Bar{V}$ satisfy the condition that the backflow matrix $\begin{bmatrix}
C_\btheta(c) & F_\btheta(c)    
\end{bmatrix}$ is of rank at least $r$, there always exists some column vector from $C_\btheta(c)$ to mix in to make $F_\btheta(c)$ full-rank. 

Moreover, as $K$ is an upper-triangular square matrix with constant entries (not dependent on $c$), and $\det K=1$, this mixing will not change the wave function amplitude of HFDS. The practical realization on HFDS for this mixing is simply to add one more linear layer with weights constructed from $K$ to the output layer of the NN that generates the backflow matrix. This new linear layer can just be absorbed into the last layer of the original NN, as a result, this mixing can be achieved by simply adjusting the weights and bias on the original output layer even without changing the NN architecture.

Since the effect of the mixing matrix $K$ is to mix new vectors into $F_\btheta(c)$, the choices for the non-zero entries of $K$ are almost generic, as long as it circumvents the situation where the mixing accidentally decreases the rank for those originally full-ranked $F_\btheta(c)$, $c\in V$. But this only happens for a zero-measure set in the parameter space defining $K$, which can be excluded from the determination of $K$.  

At last, we consider the situation where the backflow matrix has rank less than $r$, which indicates that there are not sufficient column vectors from $C_\btheta(c)$ to make $F_\btheta(c)$ full-rank, such that $F_\btheta(c)$ is always rank-deficient, i.e., $\det F_\btheta(c)=0$ no matter what mixing matrix $K$ we have chosen. In this case, not only the columns of the $r\times (N_s+r)$ backflow matrix are linearly dependent, but the rows are also linearly dependent. As a consequence, the $\rhfeq$ matrix for HFDS is rank-deficient, and will give zero amplitude from its determinant. Nevertheless, Eqn.~\eqref{eq:rhf} is still valid, since it also gives zero amplitude as the original HFDS amplitude, due to $\det F_\btheta(c)=0$. Although it brings with an ill-defined $F^{-1}$ in the expression, in practice, this none-valued output can be replaced with 0 during programming without affecting the performance.

In fact, this mixing procedure is also able to bring $F_\btheta(c)$ away from rank-deficient regimes of $\btheta$, where $F_\btheta(c)$ has extremely small but nonzero singular values and $F^{-1}$ has extremely large entries, via, for example, optimizing the entries of matrix $K$ with some proper loss function. By doing so, the learning of $\rhfeq$ with $\rjbfeq$ or $\rbfeq$ will be practically easier.

\section{Explicit Construction for Subset SPO}
{\label{append:constructSubset}}
In Sec.~\ref{sec:NNBF_HF_SPO}, we give statements on the relation between SPO's from different classes of wave function ansatz. Here we provide explicit construction of SPO to support some of these statements. 

\begin{enumerate}
\item 
We start with $\rhfeq \underset{\text{spo}}\subset  (r+1)\text{-HF}$ for any $r$. Recall that the SPO for $\rhfeq$ is given by $A-BF^{-1}C$, which is an $L_s\times N_s$ matrix; it is straightforward to see that it can be reproduced in the class of $(r+1)$-HF, by letting 
\begin{equation}
A' = A,\quad 
B' = 
\begin{bmatrix}
  B
  & \rvline & 
 \begin{matrix}
  0 \\ 0 \\ \vdots \\ 0
  \end{matrix}
\end{bmatrix}
\label{eq:constrHFr+1SPOab}
\end{equation}
\begin{equation}
C' = \begin{bmatrix}
C \\
\hline
\begin{matrix}
0 & 0 & \cdots & 0    
\end{matrix}
\end{bmatrix}, \quad
F' = \begin{bmatrix}
F & \rvline & 
\begin{matrix}
0 \\ 0 \\ \vdots \\ 0 
\end{matrix}
\\
\hline 
\begin{matrix}
 0 & 0 & \cdots & 0   
\end{matrix}
& \rvline
& 1
\end{bmatrix}
\label{eq:constrHFr+1SPOcf}
\end{equation}
such that $A'-B'F'^{-1}C'= A-BF^{-1}C$ for any $c$. Note that the constant elements on the backflow matrices $C$ and $F$ can be constructed by assigning the weights on the final linear layer to those positions as 0 and the bias as the corresponding constant values (0 or 1 here).

Similarly, we can also show that $\rjbfeq \underset{\text{spo}}\subset (r+1)\text{-JBF}$ and $\rbfeq \underset{\text{spo}}\subset (r+1)\text{-BF}$, using the same construction as $\rhfeq$ (Eqn.~\eqref{eq:constrHFr+1SPOab} and \eqref{eq:constrHFr+1SPOcf}), such that $A'-B'C'=A-BC$ for any $c$.
\item Next, we show that for any $r$, including the cases of $r\geq L_s$, we have $\rjbfeq  \underset{\text{spo}}\subset L_s\text{-JBF}$ and $\rbfeq  \underset{\text{spo}}\subset L_s\text{-BF}$, both of which amount to showing $A'-B'C'=A-BC$, where $B'$, $C'$ are now matrices of size $L_s\times L_s$ and $L_s\times N_s$, respectively. Copy the neural network generating $C$ for $\rbfeq$ or $\rjbfeq$ and add one more linear layer on top of it. The weights and bias of this additional layer are given by
\begin{equation}
W_{(mn)(kl)} = \delta_{nl}B_{mk}, \quad b_{(mn)}=0    
\label{eq:newlinear_wb}
\end{equation}
where we use the matrix index $(mn), (kl)$ to index the weights and bias. The output of this new linear layer will directly give us the $L_s\times N_s$ matrix $C'=BC$, because
\begin{equation}
\begin{split}    
C'_{mn} =& \sum_{k,l}W_{(mn)(kl)}C_{(kl)}+b_{(mn)}\\=&\sum_{k}B_{mk}C_{kn}\equiv(BC)_{mn}   
\end{split}
\end{equation}
Note that this additional linear layer $(W',b')$ can be absorbed into previous linear layer $(W,b)$ for $C$ simply by $W\leftarrow W'W$, $b\leftarrow W'b+b'$. As we have gotten $C'=BC$, this construction is completed by the assignments of $A'=A$ and $B'=I_{L_s}$. 

\item 
Here we demonstrate that, given an NNBF that outputs matrix $M$ (see Eqn.~\ref{eq:NNBF0}), we can generate an $L_s$-BF, that outputs matrix $C$, such that $BC=M$.

We start by rewriting a general $L_s$-BF where we absorb $B$ into the NN, which amounts to adding one more linear layer with weights as 
\begin{equation}
W' = \begin{bmatrix}
B & \bm{0} & \cdots & \bm{0} \\
\bm{0} & B & \cdots & \bm{0} \\
\vdots & \vdots & \ddots & \vdots \\
\bm{0} & \bm{0} & \cdots & B \\
\end{bmatrix}    
\end{equation}
where this is an $N_s\times N_s$ block matrix; We can combine it with the weight matrix from previous linear layer $W$ which is given as
\begin{equation}
W = \begin{bmatrix}
W_1 \\ W_2 \\ \vdots \\ W_{N_s}    
\end{bmatrix}    
\end{equation}
where $W_h$, $h=1,2,...,N_s$ ($H$ is the width of last hidden layer) are arbitrary $r\times H$ matrices. 
Now the resulting weight $W''=W'W$ is still a generic $(L_sN_s)\times H$ matrix with no constraints, because each block of $W''$ is given by
\begin{equation}
W''_h = BW_h    
\end{equation}
and is of full-rank $L_s$ as $B$ and $W_h$ are of size $L_s\times r$ and $r\times H$, respectively, with $r\geq L_s$.
We can assume $B$ is of full rank $L_s$ (if not, it can be brought to full rank by mixing with $A$, see Appendix.~\ref{append:hfdsInvertF}), such that there are complete sets of linearly independent vectors within it. Finally, the arbitrary choice of $W_h$ enables $W''_h$ to be any $L_s\times H$ matrix.  

The reverse direction of the relation between NNBF and $\rbfeq$ ($r\geq L_s$) is also true by simply choosing a common basis set as $B$ for each $W_h$.

\end{enumerate}

\section{Absorption of Determinant Factor into SPO}
{\label{append:JastrowAbsorbSPO}}
In this section, we show the derivation for the absorption of determinant factor into SPO for $\rhfeq$ and $\rjbfeq$.

For $r\leq N_s$, we embed the $r\times r$ matrix $F_{\btheta}(c)$ in an $N_s\times N_s$ block-diagonal matrix with identity on the extra diagonal block 
\begin{equation}
[F_\btheta] \rightarrow 
\begin{bmatrix}
F_\btheta & \rvline & \bm{0} \\
\hline 
\bm{0} & \rvline & I_{N_s-r}
\end{bmatrix}
\equiv \widetilde{F}_\btheta
\end{equation}
This leaves the wave function amplitude unchanged, since $\det F=\det \widetilde{F}$, but allows us to multiply the two square matrices from the determinant factor and from the SPO's together  before the determinants are computed.  
\begin{equation}
\begin{split}
\Psi_\rhfeq(c) &= \det F_{\bm{\theta}}(c)\det( (A - BF^{-1}_{\bm{\theta}}(c)C_{\bm{\theta}}(c))[c])\\
&=\det \widetilde{F}_{\bm{\theta}}(c) \det( (A - BF^{-1}_{\bm{\theta}}(c)C_{\bm{\theta}}(c))[c]) \\
&=\det \left[ (A-BF^{-1}C)\widetilde{F} [c]\right]\\
&=\det \left[ (A+A(\widetilde{F}-I)-BF^{-1}C\widetilde{F})[c]\right]
\end{split}    
\label{eq:rhf_detxspo}
\end{equation}
From Eqn.~\eqref{eq:rhf_detxspo}, we see a new SPO matrix of size $L_s\times N_s$. Next we transform it into BF.

After writing $C$ in terms of block matrices as well, i.e., 
\begin{equation}
 C = \begin{bmatrix}
    C_1 & \rvline & C_2 
 \end{bmatrix}   
\end{equation}
where $C_1$ and $C_2$ are of size $r\times r$ and $r\times (N_s-r)$, respectively, we have 
\begin{equation}
\begin{split}
BF^{-1}C\widetilde{F}&=BF^{-1}
\begin{bmatrix}
C_1 & C_2   
\end{bmatrix}    
\begin{bmatrix}
F & \bm{0}\\
\bm{0} & I
\end{bmatrix}\\
&=B\begin{bmatrix}
F^{-1}C_1F & F^{-1}C_2   
\end{bmatrix}
\end{split}
\label{eq:cf_transform}
\end{equation}

Meanwhile, with $A$ also in block matrix form, i.e., 
\begin{equation}
A = \begin{bmatrix}
 A_{11}  & \rvline & A_{12} \\
 \hline 
 A_{21}  & \rvline & A_{22}
\end{bmatrix} =
\begin{bmatrix}
 A_1 & \rvline & A_2   
\end{bmatrix}
\end{equation}
where $A_1$ and $A_2$ are of size $L_s\times r$ and $L_s\times(N_s-r)$, respectively, $A_{11}$ and $A_{21}$ are of size of $r\times r$ and $(L_s-r)\times r$, respectively, etc, we have 
\begin{equation}
\begin{split}
A(\widetilde{F}-I)&=
\begin{bmatrix}
A_{11} & A_{12} \\
A_{21} & A_{22}
\end{bmatrix}
\begin{bmatrix}
F-I & \bm{0} \\
\bm{0} & \bm{0}
\end{bmatrix}\\&=
\begin{bmatrix}
A_{11}(F-I) & \bm{0} \\
A_{21}(F-I) & \bm{0}
\end{bmatrix}\\&=A_{1}
\begin{bmatrix}
(F-I) & \bm{0}
\end{bmatrix}
\end{split}
\label{eq:af_transform}
\end{equation}
Combining Eqn.~\eqref{eq:cf_transform} and Eqn.~\eqref{eq:af_transform} together, we eventually obtain 
\begin{equation}
\begin{split}    
\Psi_{\text{HF}}(c) &= \det \begin{bmatrix}
A[c] & B[c] \\
C_{\bm{\theta}}(c) & F_{\bm{\theta}}(c)
\end{bmatrix} \\ &=
\det\left(  (A-B'C_{\bm{\theta}}'(c))[c] \right)
\end{split} 
\end{equation}
where 
\begin{equation}
B' = 
\begin{bmatrix}
-A_1 & B    
\end{bmatrix},
\quad
C_\btheta'(c) = \begin{bmatrix}
F-I & \bm{0} \\
F^{-1}C_1F & F^{-1}C_2
\end{bmatrix}
\end{equation}
are of size $L_s\times 2r$ and $2r\times N_s$, respectively.

Likewise, for $\rjbfeq$, we have
\begin{equation}
\begin{split}    
\Psi_{\text{JBF}}(c) &= \det\left(F_{\bm{\theta}}(c)\right)\det\left((A-BC_{\bm{\theta}}(c))[c]\right) \\ &=
\det\left(  (A-B'C_{\bm{\theta}}'(c))[c] \right)
\end{split} 
\end{equation}
where 
\begin{equation}
B' = 
\begin{bmatrix}
-A_1 & B    
\end{bmatrix},
\quad
C_\btheta'(c) = \begin{bmatrix}
F-I & \bm{0} \\
C_1F & C_2
\end{bmatrix}
\end{equation}
are of size $L_s\times 2r$ and $2r\times N_s$, respectively.

\section{Relation between Wave Function Fidelity and SPO Distance}
\label{append:fide_spo_relation}
In Fig.~\ref{fig:spo-fide}, we observe that at $r=1,2$, though the SPO distance Eqn.~\eqref{eq:spoloss} is small (Fig.~\pref{fig:spo-fide}{a}), the wave function fidelity Eqn.~\eqref{eq:fidewf} is relatively worse (Fig.~\pref{fig:spo-fide}{b}), compared with larger $r$ cases. In this section we will look into this observation by showing the relation between the wave function fidelity and the SPO distance.  

Within VMC, the wave function fidelity can be calculated as
\begin{equation}
\mathcal{F}_{\text{wf}} = \frac{\left(\frac1S\sum_{\{c\}}\frac{\Phi(c)}{\Psi(c)}\right)^2}{\frac1S\sum_{\{c\}}\left(\frac{\Phi(c)}{\Psi(c)}\right)^2}    
\end{equation}
where $S$ is the size of sample set $\{c\}$ sampled from $|\Psi(c)|^2$. As $\Phi(c)/\Psi(c)$ is close to 1 here, we rewrite it as 
\begin{equation}
\frac{\Phi(c)}{\Psi(c)}\equiv 1+x(c)    
\end{equation}
For most of $c$, we should have $|x(c)|\ll 1$. We relate $x(c)$ to the the SPO distance component with Taylor expansion around $\Psi(c)$ up to 1st order, 
\begin{equation}
x(c) = \frac{\det\bar{A}(c)}{\det\bar{A}_0(c)}-1\approx \sum_{jk} \bar{A}_0(c)^{-1}_{kj}\left(\bar{A}(c)-\bar{A}_0(c)\right)_{jk}   
\end{equation}
where $\bar{A}(c)$ and $\bar{A}_0(c)$ are the square matrices in the determinant wave function form of $\ket{\Phi}$ and $\ket{\Psi}$, respectively.    
Note that the truncation in Taylor expansion is one of the sources of error for our final approximation formula, which can nonetheless be systemically improved by including higher-order terms.

\begin{figure}[h]
\centering
\begin{subfigure}[b]{0.48\textwidth}
\centering
\includegraphics[width=\textwidth]{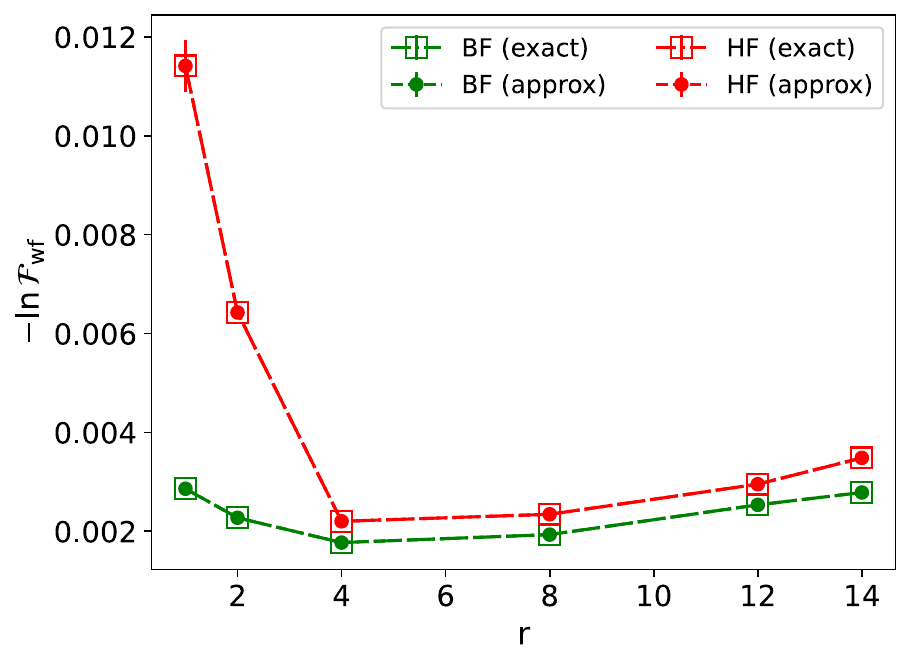}
\end{subfigure}
\caption{Comparison between the approximate results Eqn.~\eqref{eq:approxFwf} for the wave function fidelity and the exact results from VMC. The target wave functions are $\rjbfeq$ from ground state approximation, and $2r$-BF and $\rhfeq$ are optimized towards the target wave functions by minimizing the SPO distances.}
\label{fig:Spo-Fide-relation}
\end{figure}

In the following, we will perform Taylor expansion up to 2nd order on wave function fidelity as well. The first-order term will vanish as $\mathcal{F}_{\text{wf}}=1$ is a stationary point. After defining 
\begin{equation}
\overline{x}=\frac1S\sum_{\{c\}} x(c),\quad \overline{x^2}=\frac1S\sum_{\{c\}} x(c)^2     
\end{equation}
and $|\overline{x}|\ll1$, $|\overline{x^2}|\ll1$, we can either use chain rule to obtain the Hessian as
\begin{equation}
\frac{\delta}{\delta \bar{A}(c')_{j'k'}}\frac{\delta\ln\mathcal{F}_{\text{wf}}}{\delta \bar{A}(c)_{jk}}=\frac{1}{S^2} \bar{A}_0(c)^{-1}_{kj} \bar{A}_0(c')^{-1}_{k'j'}(1-S\delta_{c,c'})    
\end{equation}
or directly expand on $\ln\mathcal{F}_{\text{wf}}$ in terms of $\overline{x}$ and $\overline{x^2}$, we obtain the approximation formula for $\ln\mathcal{F}_{\text{wf}}$ , the deviation from perfect fidelity $\mathcal{F}_{\text{wf}}=1$, as the variance of $x(c)$,
\begin{equation}
\ln\mathcal{F}_{\text{wf}}\approx \overline{x}^2-\overline{x^2}    
\label{eq:approxFwf}
\end{equation}
In Fig.~\ref{fig:Spo-Fide-relation}, we examine the accuracy of Eqn.~\eqref{eq:approxFwf} by comparing with the exact results from VMC.

\begin{figure*}
\centering
\includegraphics[width=\textwidth]{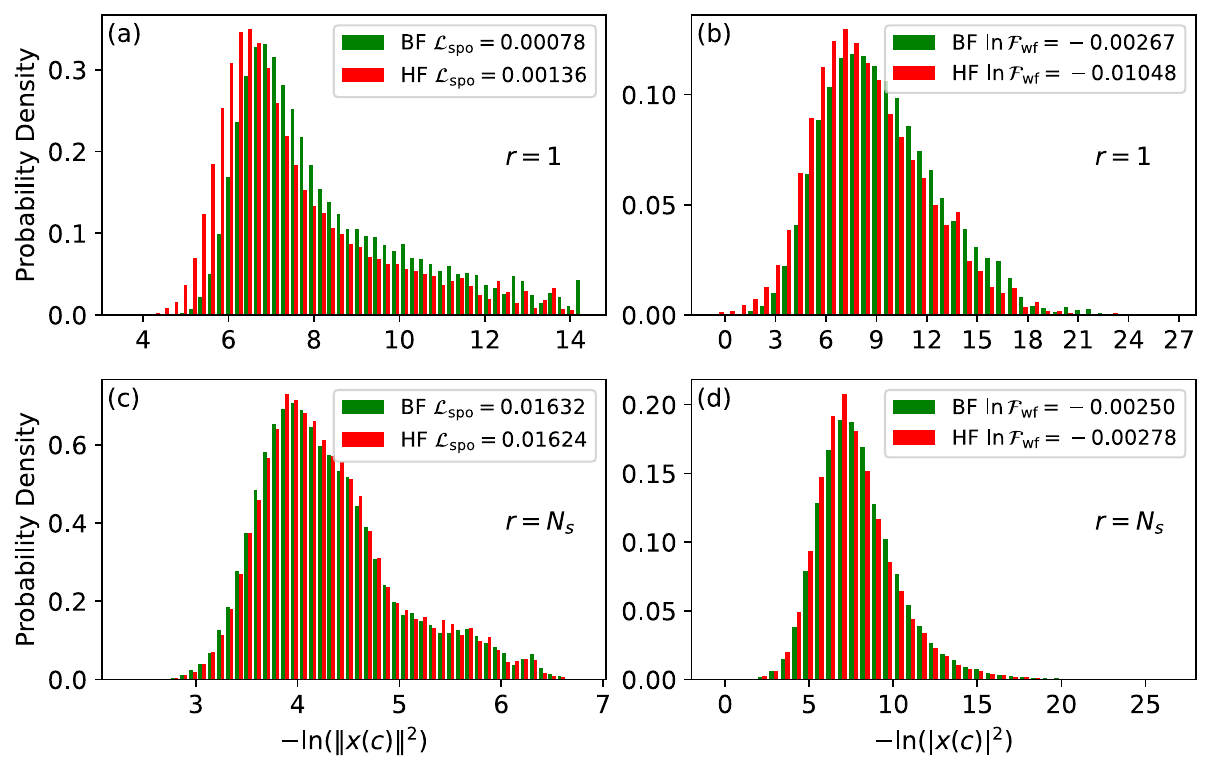}
\caption{Probability distributions of $\Vert x(c)\Vert^2$ and $\vert x(c)\vert^2$ at $r=1$ and $14(N_s)$. The corresponding SPO distance and wave function fidelity are given in the legends.}
\label{fig:Spo-Fide-relation2}
\end{figure*}

We also found that $\overline{x}^2\ll \overline{x^2}$. Then by just comparing $\overline{x^2}$ with the SPO distance Eqn.~\eqref{eq:spoloss}, which we rewrite as
\begin{equation}
\mathcal{L}_{\text{spo}}=\frac1S\sum_{\{c\}}\Vert x(c) \Vert^2 = \frac1S\sum_{\{c\}}\sum_{jk}\vert \bar{A}(c)-\bar{A}_0(c) \vert_{jk}^2   
\end{equation}
we can see that wave function fidelity is effectively a weighted average of SPO distance on each element, with the weights given by $\Bar{A}_0(c)^{-1}$ from the target state. Further evidence is shown from the comparison between Fig.~\pref{fig:Spo-Fide-relation2}{a,c} and Fig.~\pref{fig:Spo-Fide-relation2}{b,d}, where the difference of wave function fidelity is affected not only by the SPO distance, but also the target wave function.   

\section{Reconstruction of $r$-JBF with $2r$-BF}
{\label{append:rjbf_from_2r_bf}}
Despite the existence of matrix multiplication within $\rjbfeq$, it is possible to reconstruct $\rjbfeq$ rigorously from $2r$-BF with feed-forward neural networks still. In this section we show it explicitly.  

First, with the feed-forward neural network that generates backflow matrices $C_\btheta(c)$ and $F_\btheta(c)$ for $\rjbfeq$, we add an activation function $a(x)=\ln(x)$ to this layer (note that originally for $\rjbfeq$, this layer does not have activation function on it. Also note that for the case where $x=0$, we can nonetheless replace it with $x+\epsilon$, $\epsilon\ll1$, which will not affect the subsequent steps). Now, we have the output from this modified layer as $\ln C_{jk}$ $(j=1,2,...,r$; $k=1,2,...,N_s)$, and $\ln F_{lm}$ $(l,m=1,2,...,r)$. In the following, we will work on the first $r$ columns of $\ln C$, i.e., $\ln C_1$, in order to reproduce $C_1F$ with feed-forward neural network. As for the remaining $\ln C_2$, we construct the weights and bias as an identity map, and the non-linear activation function introduced later will bring $\ln C_2$ back to the original $C_2$ from \rjbfeq.  

Next, we add one more layer to this neural network, whose activation function is $a(x)=\exp(x)$, bias are 0's, and weights are
\begin{equation}
W^{[C]}_{(xyz)(jk)}=\delta_{xj}\delta_{zk}, \quad W^{[F]}_{(xyz)(lm)}=\delta_{zl}\delta_{ym}    
\end{equation}
where $x,y,z=1,2,...,r$, such that the output from this layer is 
\begin{equation}
\begin{split}    
D_{xyz}=&\exp\left(\sum_{jk}W^{[C]}_{(xyz)(jk)}\ln C_{1,jk}+\sum_{lm}W^{[F]}_{(xyz)(lm)}\ln F_{lm}\right)  \\  
=& C_{1,xz}F_{zy}
\end{split}
\end{equation}

At the end, we add one more linear layer with non-zero weights as 
\begin{equation}
W_{(jm)(xyz)}=\delta_{jx}\delta_{my}    
\end{equation}
such that the output is 
\begin{equation}
\widetilde{C}_1=\sum_{xyz}W_{(jm)(xyz)}D_{xyz}=\sum_z C_{1,jz}F_{zm}\equiv (C_1F)_{jm}   
\end{equation}

Overall, by introducing two more layers of widths $r^2(N_s+r)$ and $r(N_s+r)$  to the original neural network for \rjbfeq, we are able to reproduce the backflow matrices with $2r$-BF. Therefore, we have analytically proved that \rjbfeq~is a subset of \rbfeq  at finite $w$ if we allow $\rbfeq$ to have additional layers. 

\section{Computational Cost}
{\label{append:compute_cost}}
In this section, we give the computational cost for obtaining the amplitude from various types of NNBF wave functions.

Let us start with NN part. The input size is $L_s$, and the output size is $rN_s+r^2$, and assume the width of NN is $w$, then the computational cost due to the matrix-vector multiplication is given by $O(w^2+wL_s+wr^2+wrN_s)$. 

Next, the cost of the evaluation of determinant is
\begin{equation}
\begin{split}
r\text{-HF}: \quad &  O((N_s+r)^3)=O(N_s^3+r^3) \\
r\text{-BF}: \quad & O(rN_s^2+N_s^3) \\
r\text{-JBF}: \quad & O(r^3+rN_s^2+N_s^3)
\end{split}   
\end{equation}

In Appendix.~\ref{append:rjbf_from_2r_bf}, we show that to reproduce \rjbfeq~from $2r$-BF, we can add two more layers of width $r^3+r^2N_s$ and $rN_s+r^2$, this will give an extra cost as $O(r^3(N_s+r)^2)=O(r^3N_s^2+r^5)$.

\section{Failure of SPO Matching with JSD towards NNBF}
{\label{append:spo_jsd_nnbf}}
In this section, we illustrate the impossibility of matching the SPO of NNBF with JSD for rank $r\leq N_s$, thus, combining with the numerical evidences Fig.~\pref{fig:Energy-Rank}{b} and Fig.~\ref{fig:unif_spo}, we conclude that NNBF represents a broader space of SPO. 

For simplicity, we consider the case of $N_s$-JSD and $N_s$-BF. Since the NN generating the back flow matrices has linear layer as the last layer, there are constant matrices $b$ in the backflow matrices, then we rewrite the SPO as
\begin{equation}
\begin{split}
\mathcal{S}_{\text{JSD}}(c)&=A_{\text{JSD}}F_{\text{JSD},\btheta}(c)\\
&=A_{\text{JSD}}b_{\text{JSD}}+A_\text{JSD}\widetilde{F}_{\text{JSD},\btheta}(c) \\     
\mathcal{S}_{\text{BF}}(c)&=A_\text{BF}-B_\text{BF}C_{\text{BF},\btheta}(c)\\
&=A_\text{BF}-B_\text{BF}b_\text{BF}-B_\text{BF}\widetilde{C}_{\text{BF},\btheta}(c)
\end{split}   
\end{equation}
In order to match the SPO, $\mathcal{S}_{\text{JSD}}(c)=\mathcal{S}_{\text{BF}}(c)$, we have to solve the following equations for static matrices
\begin{equation}
\begin{split}
A_\text{JSD}&=-B_\text{BF}\\
A_{\text{JSD}}b_{\text{JSD}}&=A_\text{BF}-B_\text{BF}b_\text{BF}
\end{split}    
\end{equation}
However, the second equation above is an over-determinant equation, so there is no exact solution for $b_{\text{JSD}}$, unless $A_\text{BF}-B_\text{BF}b_\text{BF}$ falls into the linear space spanned by the column vectors of $A_{\text{JSD}}$. But this special case is not generically true for the BF wave function we are considering. 

In conclusion, we have showed that it is impossible for JSD to reproduce the same SPO from BF.

\section{Supervised Wave-function Optimization}

{\label{append:swo}}
\begin{figure*}[t]
\centering
\includegraphics[width=\textwidth]{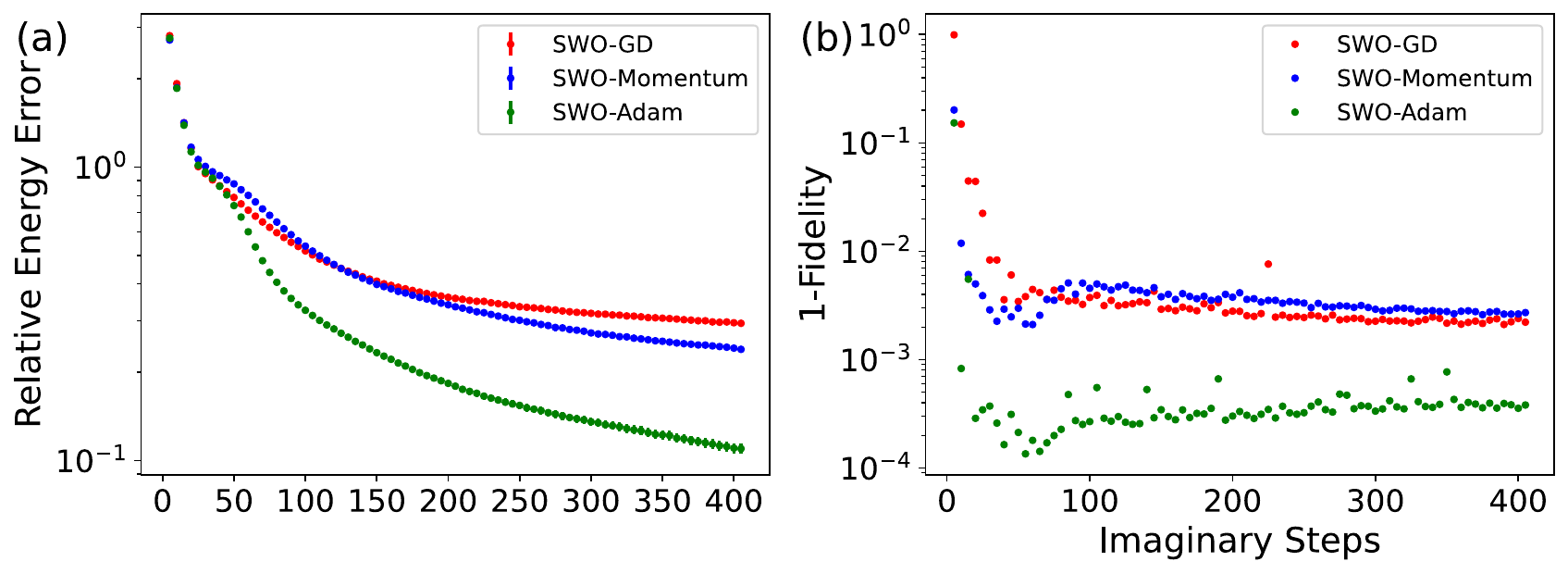}
\caption{(a) Energy and (b) fidelity (associated with the loss function Eqn.~\eqref{eq:LossFun}) from $\rhfeq$ at $r=64$ during SWO optimization ($\tau=0.018$) with optimizer SGD (red), Momentum (blue) and Adam (green) for Hubbard model ($L=4\times8$, PBC, $U/t=8$, $N_\uparrow=N_\downarrow=14$).}
\label{fig:SWO-Opt}
\end{figure*}

In this section, we provide detailed information on the SWO method, including the derivation for the gradient formula Eqn.~\eqref{eq:GradEst} we use for NNBF (and its variants) and comparison between various optimizers.

To begin with, we note that SWO is a first-order optimization method for variational wave function to undergo imaginary time evolution in a stochastic way. As already mentioned in Sec.~\ref{sec:method}, at each imaginary-time step $t$, we set the target state as the state evolved by imaginary time $\tau$, i.e, $\ket{\Phi_t}=e^{-\tau\hat{H}}\ket{\Psi_{t-1}}\approx (1-\tau\hat{H})\ket{\Psi_{t-1}}$, where $\tau$ needs to be small enough for the latter approximation to be valid. In practice, we notice that the energy can be minimized successfully with $\tau<1/E_{t=0}$, where $E_{t=0}$ is the variational energy of randomly initialized state.

To derive the gradient formula Eqn.~\eqref{eq:GradEst}, we first take derivative of the loss function Eqn.~\eqref{eq:LossFun} with respect to the parameter $\btheta$,
\begin{equation}
\begin{split}
\frac{\partial\mathcal{L}}{\partial\btheta} &= -\frac{\partial_{\btheta}\braket{\Psi_{t}}{\Phi_{t}}}{\braket{\Psi_{t}}{\Phi_{t}}}-\frac{\partial_{\btheta}\braket{\Phi_{t}}{\Psi_{t}}}{\braket{\Phi_{t}}{\Psi_{t}}}+\frac{\partial_{\btheta}\braket{\Psi_{t}}{\Psi_{t}}}{\braket{\Psi_{t}}{\Psi_{t}}} \\
&=-\frac{\sum_c(\partial_\btheta\braket{\Psi_t}{c})\braket{c}{\Phi_t}}{\sum_c\braket{\Psi_t}{c}\braket{c}{\Phi_t}}-\frac{\sum_c\braket{\Phi_t}{c}(\partial_\btheta\braket{c}{\Psi_t})}{\sum_c\braket{\Phi_t}{c}\braket{c}{\Psi_t}}\\
&+\frac{\sum_c(\partial_\btheta\braket{\Psi_t}{c})\braket{c}{\Psi_t}+\sum_c\braket{\Psi_t}{c}(\partial_\btheta\braket{c}{\Psi_t})}{\sum_c\braket{\Psi_t}{c}\braket{c}{\Psi_t}}
\end{split}
\label{eq:derivLoss}
\end{equation}
This expression is exact since we are summing over all the configurations $c$. Next, we use the sample set $\{c|p(c)\}$ from probability distribution $p(c)=|\Psi_{t-1}(c)|^2$ to replace the exact summation. Then, for example, the first term of Eqn.~\eqref{eq:derivLoss} ends up as
\begin{equation}
\begin{split}
&\frac{\sum_c(\partial_\btheta\braket{\Psi_t}{c})\braket{c}{\Phi_t}}{\sum_c\braket{\Psi_t}{c}\braket{c}{\Phi_t}}=\frac{\sum_c\frac{(\partial_\btheta\braket{\Psi_t}{c})\braket{c}{\Phi_t}}{p(c)}p(c)}{\sum_c\frac{\braket{\Psi_t}{c}\braket{c}{\Phi_t}}{p(c)}p(c)} \\
\approx & \frac{\sum_{\{c\}}\frac{(\partial_\btheta\braket{\Psi_t}{c})\braket{c}{\Phi_t}}{p(c)}}{\sum_{\{c\}}\frac{\braket{\Psi_t}{c}\braket{c}{\Phi_t}}{p(c)}}
=\frac{\sum_{\{c\}}\frac{(\partial_\btheta\braket{\Psi_t}{c})\braket{c}{\Phi_t}}{\braket{\Psi_{t-1}}{c}\braket{c}{\Psi_{t-1}}}}{\sum_{\{c\}}\frac{\braket{\Psi_t}{c}\braket{c}{\Phi_t}}{\braket{\Psi_{t-1}}{c}\braket{c}{\Psi_{t-1}}}}
\end{split}
\label{eq:GradVMC}
\end{equation}

As we are optimizing over determinant wave function form, i.e., $\Psi_\btheta(c)=\det\widetilde{A}_\btheta(c)$, we can further simplify the derivative term in Eqn.~\eqref{eq:derivLoss} as 
\begin{equation}
\begin{split}    
\frac{\partial\Psi(c)}{\partial\btheta}&=\frac{\partial\det\widetilde{A}_\btheta(c)}{\partial\btheta}\\
&=\det\widetilde{A}_\btheta(c)\Tr\left(\widetilde{A}_\btheta(c)^{-1}\frac{\partial \widetilde{A}_\btheta(c)}{\partial\btheta}\right)\\
&=\Psi(c)\Tr\left(\widetilde{A}_\btheta(c)^{-1}\frac{\partial \widetilde{A}_\btheta(c)}{\partial\btheta}\right)
\end{split}
\end{equation}
After defining the amplitude ratio $\phi(c)$ and $\psi(c)$ as Eqn.~\eqref{eq:ampratio} to simplify the notations, we factor out the derivative part, and put together all the other factors as a coefficient $\alpha(c)$ given by Eqn.\eqref{eq:alphacoe}, then we end up with the gradient estimation formula Eqn.~\eqref{eq:GradEst}. That only the real part of gradient is taken is because the parameter set $\btheta$ are real numbers and 
each term in Eqn.~\eqref{eq:derivLoss} is accompanied with its complex conjugate.

Finally, we note that, as SWO is a first-order optimization method, it can be further improved with ML optimizers, such as Momentum-based gradient descent, Adam, etc. In Fig.~\ref{fig:SWO-Opt}, we compare the performance from gradient descent (GD), momentum and Adam within SWO on the ground state approximation with $\rhfeq$. It turns out that SWO-Adam is more efficient than other two optimizers, as it optimizes the fidelity to a higher value at each step, so as to minimize the energy faster than the other two optimizers. Henceforth, we use SWO-Adam for optimization in our numerical tests.

\section{Notations}
In Table~\ref{tab:table1}, we give a summary for the notations used in this paper.
\begin{table}[t]
\caption{\label{tab:table1}
Table of Notations}
\begin{ruledtabular}
\begin{tabular}{lcdr}
$c$ & $L_s$-size binary vector \\
$\btheta$ & Parameters (weights and bias) of neural networks \\
$A$ & $L_s\times N_s$ constant matrix \\
$A[c]$ & $N_s\times N_s$ matrix from row selection on $A$ based on $c$\\
$B$ & $L_s\times r$ constant matrix \\
$B[c]$ & $N_s\times r$ matrix from row selection on $B$ based on $c$\\
$C_{\btheta}(c)$ & $r\times N_s$ 
backflow matrix from neural networks, \\
$F_{\btheta}(c)$ & $r\times r$ 
backflow matrix from neural networks, \\

$\rhfeq$ & $\Psi(c)=\det(F_\btheta(c))\det\left((A-BF_{\btheta}(c)^{-1}C_\btheta(c))[c]\right)$   \\

$\rjbfeq$ & $\Psi(c)=\det(F_\btheta(c))\det\left((A-BC_\btheta(c))[c]\right)$   \\

$\rbfeq$ & $\Psi(c)=\det\left((A-BC_\btheta(c))[c]\right)$   \\

$\rjnnbfeq$ & $\Psi(c)=\det(F_\btheta(c))\Psi_{L_s\text{-BF}}(c)$   \\

$r$-JSD & $\Psi(c)=\det(F_\btheta(c))\det A[c]$   \\

$r$-expJBF & $\Psi(c)=\exp\left({f_{\btheta}(c)}\right)\Psi_{r\text{-BF}}(c)$   \\

$r$-expJsBF & $\Psi(c)=\exp\left({f_{1,\btheta}(c)}\right)\cos\left({f_{2,\btheta}(c)}\right)\Psi_{r\text{-BF}}(c)$   \\

\end{tabular}
\end{ruledtabular}
\end{table}


\bibliography{main}
\end{document}